\documentclass[twocolumn,prd,showpacs]{revtex4}
\usepackage{graphicx}
\begin{document}

\newcommand{\re}{\mathop{\mathrm{Re}}}

\newcommand{\be}{\begin{equation}}
\newcommand{\ee}{\end{equation}}
\newcommand{\bea}{\begin{eqnarray}}
\newcommand{\eea}{\end{eqnarray}}


\title{Brane universes tested against astronomical data}

\author{Mariusz P. D\c{a}browski}
\email{mpdabfz@uoo.univ.szczecin.pl}
\affiliation{\it Institute of Physics, University of Szczecin, Wielkopolska 15,
          70-451 Szczecin, Poland}
\author{W{\l }odzimierz God{\l }owski}
\email{godlows@oa.uj.edu.pl}
\affiliation{\it Astronomical Observatory, Jagiellonian University, 30-244
Krakow, ul. Orla 171, Poland}
\author{Marek Szyd{\l }owski}
\email{uoszydlo@cyf-kr.edu.pl}
\affiliation{\it Astronomical Observatory, Jagiellonian University, 30-244
Krakow, ul. Orla 171, Poland}

\date{\today}

\input epsf

\begin{abstract}
We discuss observational constrains coming from supernovae
imposed on the behaviour of the Randall-Sundrum models.
We test the models using the
Perlmutter SN Ia data as well as the new Knop and Tonry/Barris samples.
The data indicates that, under the assumption that we admit zero pressure dust matter
on the brane, the cosmological constant is still needed
to explain current observations.
We estimate the  model parameters using the best-fitting procedure and
the likelihood method.
The observations from supernovae give a large value of the density
parameter for brane matter $\Omega_{\lambda,0} \simeq 0.01$ as the best fit.
For high redshifts $z > 1.2$, the difference between the brane model and
the $\Lambda$CDM (Perlmutter) model becomes detectable observationally.
From the maximum likelihood method we obtained the favored value of
$\Omega_{\lambda,0} = 0.004 \pm 0.016$ for $\Omega_{k,0}=0$ and
$\Omega_{\mathrm{m},0}=0.3$. This gives the limit $\Omega_{\lambda,0} < 0.02$
at $1 \sigma$ level. While the model with brane effects is preferred
by the supernovae type Ia data, the model without brane fluid is still
statistically admissible.
We also discuss how fit depends on restrictions of the sample, especially
with respect to redshift criteria. We also pointed out the property of
sensitive dependence of results with respect to choice of ${\cal M}$ parameter.
For comparison the limit on brane effects which comes from CMB anisotropies and
BBN is also obtained. The uncertainty in the location of the first peak
gives a stronger limit $\Omega_{\lambda,0} < 1.0 \cdot 10^{-12}$, whereas from BBN
we obtain that $\Omega_{\lambda,0} < 1.0 \cdot 10^{-27}$. However, both very strict
limits are obtained with the assumption that
brane effects do not change the physics in the pre-recombination era,
while the SN Ia limit is model independent.

We demonstrate that the fit to supernovae data can also be
obtained if we admit the phantom matter $p = - (4/3) \varrho$ on
the brane, where this matter mimics the influence of the cosmological constant.
We show that phantom matter enlarges the age of the universe on the brane which is
demanded in cosmology. Finally, we propose to check for dark radiation
and brane tension by the application of the angular diameter of galaxies minimum value test.

\end{abstract}

\maketitle


\section{Introduction}

The idea of brane universes has originated from Ho\v{r}ava and
Witten \cite{hw} followed by Randall and Sundrum \cite{rs1+2}.
In the Randall-Sundrum scenarios of the brane-world cosmology the large
extra dimensions can solve the mass hierarchy problem of the
standard model. In this model the basic cosmological equations are modified by the presence
of some extra terms which are derived from the fact that our universe is
treated as a three-brane - to which all the gauge interactions are confined
- embedded in a five-dimensional, anti-de Sitter space, in (the whole
of) which only gravity can propagate.
The cosmological evolution of such brane universes has
been extensively investigated \cite{rs1+2,Shiromizu00,Dick01,Dvali00,Deffayet01}
(for a review see \cite{Randall02}). For example, the issues related
to the cosmological constant problem \cite{Tye01}, the
cosmological perturbations \cite{Koyama97,Kodama00}, inflationary
solutions \cite{Dvali01}, homogeneity, and flatness problems
\cite{Brandenberger02,Deffayet02} were discussed.

Brane models admit new parameters which are not present in
standard cosmology (brane tension $\lambda$ and dark radiation ${\cal U}$).
From the astronomical observations of
supernovae Ia \cite{Garnavich98,Perlmutter97,Riess98,Schmidt98,Perlmutter98,Perlmutter99} one knows that the universe
is now accelerating and the best-fit model is for the 4-dimensional cosmological constant
density parameter
$\Omega_{\Lambda_4,0}=0.72$ and for the dust density parameter
$\Omega_{m,0}=0.28$ (index "0" refers to the present moment of time).
In other words, only the exotic (negative pressure) matter in
standard cosmology can lead to this global effect. On the other hand, in
brane models the $\varrho^2$ quadratic contribution in the energy density
$\varrho$ even for a small negative pressure, contributes effectively as the
positive pressure, and makes brane
models less accelerating. In this paper we argue that in order to avoid this problem one
requires much stronger negative pressure $p < - \varrho$ (phantom) matter
to be present on the brane (cf. Ref. \cite{darkenergy,phantcos}).

We concentrate on some observational constraints on the $\varrho^2$ term,
which depending on the type of matter present scales
with the cosmic scale factor as $a^{-6\gamma}$, where $\gamma$ is the barotropic
index in the equation of state $p = (\gamma -1)\varrho$ ($p$ - the pressure and
$\varrho$ - the energy density). In fact, the $\varrho^2$ term
scales as $a^{-6}$ (similarly as stiff-fluid in standard
cosmology) for dust $\gamma=1$ matter, as $a^{-8}$ for radiation $\gamma=4/3$
and as $a^2$ for phantom matter $\gamma=-1/3$.

In this paper we discuss the observational constraints imposed on the
brane-world cosmologies by supernovae type Ia observations.
Preliminary results of such analysis were made in \cite{Godlowski04}.
Since we do not know how the exotic physics of brane theory works in the early
stages of the Universe (for example,  during structure formation), then
all estimations of the brane influence onto the Cosmic Microwave
Background (CMB) and Big Bang Nucleosynthesis (BBN) are based on the assumption that this
exotic physics does not change physical processes afterwards.
The advantage of SNIa data is its independence
of the physical processes in the early universe. Therefore, despite the fact that
weaker limits from SNIa observations can be obtained, they are in this sense more valuable.
We demonstrate that
the brane model fits the SN Ia data very well. We also show that
these observations indicate a substantial contribution from the brane,
although the observational constraints from BBN and CMB restrict this contribution
to a much higher degree. However, to derive the limits from BBN and CMB
we assume that brane effects
did not change physics before the recombination epoch.
On the other hand, the less restrictive limit we obtained from SN Ia data
is not so strongly dependent on the assumptions of a model.

We argue that the admission of the brane effects for the dust
matter on the brane does not fit supernovae data without the
presence of the cosmological constant, but we can avoid this
problem if we admit phantom matter on the brane. Also, we
argue that in the near future, when new supernovae data is available
(SNAP3), the hypothesis of important brane effects at present
($\Omega_{\lambda,0}>0$) will be tested.

In Section II we present the basic equations of brane cosmology
which are studied dynamically in Section III. Sections IV and V
are devoted to a redshift-magnitude relation for brane universes
and its comparison with supernovae type Ia data. In Section VI we
discuss the angular diameter size of a galaxy for brane
universes together with the age of the universe problem. In
Section VII we obtain the restrictions onto the brane universes
from the observations of the Doppler peaks in CMB while in Section
VIII we give the restrictions which come from BBN. In Section IX
we give our conclusions.

\section{Basic equations of brane cosmology}

Ho\v{r}ava and Witten \cite{hw} and Randall and Sundrum \cite{rs1+2} provided
us with an alternative mechanism of compactification
of extra dimensions which is different from the conventional
Kaluza-Klein scheme because extra dimensions are segments (called orbifolds) rather
than compact spaces like the circle. Based on the Randall-Sundrum brane model we can
derive the effective 4-dimensional Einstein equations by projecting
5-dimensional metric onto the brane world-volume.
The dynamics of brane models is obtained from the 5-dimensional
Einstein equations with brane location at $y=0$ as \cite{Shiromizu00}
\begin{equation}
\label{eq:1}
{}^{(5)}\tilde{G}_{\mu \nu} = \kappa_{5}^{2}
[- \Lambda_{5}^{(5)} g_{\mu \nu} + \delta(y)
(-\lambda {}^{(4)}h_{\mu \nu} + {}^{(4)}T_{\mu \nu})]~,
\end{equation}
where ${}^{(5)}\tilde{G}_{\mu \nu}$ is the 5-dimensional Einstein tensor,
$^{(5)} g_{\mu\nu}$ is the 5-dimensional metric, $y$ is an
orbifold coordinate,
$\Lambda_{5}$ is the 5-dimensional cosmological constant,
${}^{(4)}h_{\mu \nu} = {}^{(5)}g_{\mu \nu} - n_{\mu} n_{\nu}$
is 4-dimensional induced metric, $n^{\alpha}$ is the unit vector normal
to the brane, ${}^{(4)}T_{\mu \nu}$ is the 4-dimensional energy-momentum
tensor, $\lambda$ is the brane tension, and
$\kappa_{5}^{2} = 8 \pi {}^{(5)}G_{N} = 8\pi/{}^{(5)}M_{p}^{3}$ is the 5-dimensional
Einstein constant.
The main point is that the 5-dimensional Planck mass
can be much less from the 4-dimensional Planck mass
${}^{(5)}M_{p} \ll {}^{(4)}M_{p} = 1.2 \cdot 10^{19} \mathrm{GeV}$
and so the electroweak scale ${}^{(5)}M_{p} \sim \mathrm{TeV}$
may be reached in the Earth accelerators.

The 4-dimensional Einstein equations on the brane are
\begin{equation}
\label{eq:2}
{}^{(5)}G_{\mu \nu} = - \Lambda_{4}^{(4)} h_{\mu \nu} +
\kappa_{4}^{2} + \kappa_{5}^{2} \Pi_{\mu \nu} - E_{\mu \nu}~,
\end{equation}
where
\begin{eqnarray}
\kappa_{4}^{2} &=& 8 \pi {}^{(4)}G_{N} = \frac{8\pi}{{}^{(4)}M_{p}^{2}}
= \kappa_{5}^{4} \frac{\lambda}{6} ~,\\
\Lambda_{4} &=& \frac{1}{2} \kappa_{5}^{2}
\left( \Lambda_{5} + \frac{1}{6} \kappa_{5}^{2} \lambda^{2} \right) ~,\\
\Pi_{\mu \nu} &=& \frac{1}{12} T T_{\mu \nu}
- \frac{1}{4} T_{\mu \alpha} T_{\nu}^{\alpha}
+ \frac{1}{24} g_{\mu \nu} [3T_{\alpha \beta}T^{\alpha \beta} -
T^{2}]~,
\end{eqnarray}
where $\Lambda_{4}$ the
4-dimensional cosmological constant and  $\kappa_{(4)}^2 = 8\pi G = 8\pi \times 10^{-38}
(GeV)^{-2}$ is the 4-dimensional Einstein constant.
For the homogeneous and isotropic Friedmann models the Einstein field equations
reduce to a generalized Raychaudhuri equation \cite{Shiromizu00}
\begin{equation}
\label{eq:7}
\dot{H} = - H^{2} - \frac{1}{6} \kappa_{4}^{2}
\left[ \rho + 3p + (2\rho + 3p) \frac{\rho}{\lambda} \right]
+ \frac{\Lambda_{4}}{4} - \frac{2{\cal U}}{\lambda \kappa_{4}^{2}}
\end{equation}
where the dark radiation ${\cal U}$ results from the Weyl tensor contribution to 5-dimensional
geometry which is related to the fact of non-vanishing of the energy-momentum tensor of matter
on the brane. The dark radiation obeys the conservation law
\begin{equation}
\label{eq:8}
\dot{{\cal U}} = -4H{\cal U}~.
\end{equation}
The conservation law for matter in the form of perfect fluid is
\begin{equation}
\label{eq:9}
\dot{\rho} = - 3(\rho + p) H~,
\end{equation}
which for barotropic fluid gives
\begin{equation}
\label{eq:10}
\dot{\rho} = - 3\gamma H \rho~,
\end{equation}
and shows that the conservation law for equation~(\ref{eq:8}) is
simply the conservation law for radiation $(\gamma = 4/3)$, except ${\cal U}$ can
take on both positive and negative values.

In Ref. \cite{Szydlo02} we gave the formalism to express dynamical equations in
terms of dimensionless observational density parameters $\Omega$.
Following Refs. \cite{Dabrowski96,AJI+II,AJIII}
we introduce the notation useful for this purpose. In this
notation the Friedmann equation for brane universes takes the form
which is the first integral of Equations (\ref{eq:7})--(\ref{eq:9})
\begin{equation}
\label{Friedroro}
H^{2} = \frac{\kappa_{(4)}^{2}}{3} \rho + \frac{\kappa_{(4)}^{2}}{6\lambda}
\rho^{2} - \frac{k}{a^{2}} + \frac{\Lambda_{4}}{3}
+ \frac{2{\cal U}}{\lambda \kappa_{(4)}^{2}} ,
\end{equation}
where $a(t)$ is the scale factor, $k=0,\pm1$ the curvature index.
After imposing conservation law (\ref{eq:10})
we have
\be
\label{conslaw}
\varrho a^{3\gamma} = {\rm const.}
\ee
which brings the Eq. (\ref{Friedroro}) to the form
\begin{equation}
\label{FriedCCC}
\frac{1}{a^2} \left( \frac{da}{dt} \right)^2 =
\frac{C_{GR}}{a^{3\gamma}} + \frac{C_{\lambda}}{a^{6\gamma}} -
\frac{k}{a^2} + \frac{\Lambda_{4}}{3} + \frac{C_{\cal U}}{a^4} .
\end{equation}
In Eq. (\ref{FriedCCC}) we have defined the appropriate constants
$C_{GR} = (1/3)\kappa_{(4)}^2a^{3\gamma} \varrho$,
$C_{\lambda} = 1/6\lambda \cdot \kappa_{(4)}^2 a^{6\gamma}
\varrho^2$, $C_{\cal U} = 2/\kappa_{(4)}^2 \lambda \cdot a^4 {\cal U}$ ,
and $C_{GR}$ is a of general relativistic nature, $C_{\lambda}$ comes as contribution
from brane tension $\lambda$, and $C_{\cal U}$ as a contribution from dark radiation.
Though in Refs. \cite{sopuerta,Szydlo02} the Eq. (\ref{FriedCCC}) was studied using
qualitative methods we briefly discuss here the cases $\gamma = 0$
(cosmological constant), $\gamma = 1/3$ (domain walls)
and $\gamma = 2/3$ (cosmic strings) which can exactly be integrable in terms of elementary
or elliptic \cite{Dabrowski96} functions.

The first case $\gamma=0$ is the
easiest, since the first two terms on the right-hand-side of
(\ref{FriedCCC}) play the role of cosmological constants similar
to $\Lambda_{4}$
\begin{equation}
\label{Friedgam0}
\frac{1}{a^2} \left( \frac{da}{dt} \right)^2 =
\left( C_{GR} + C_{\lambda} + \frac{\Lambda_{4}}{3} \right) -
\frac{k}{a^2}  + \frac{C_{\cal U}}{a^4} .
\end{equation}
The next two cases involve terms which were already integrated
in the context of general relativity. For $\gamma = 1/3$ (domain walls on the
brane) the general relativistic term $C_{GR}$ in (\ref{FriedCCC}) scales as domain walls
in general relativity while the term with $C_{\lambda}$ scales as cosmic strings (curvature)
in general relativity, i.e.,
\be
\label{Friedgam13}
\frac{1}{a^2} \left( \frac{da}{dt} \right)^2 =
 \frac{C_{\lambda}-k}{a^{2}} + \frac{C_{GR}}{a}
+ \frac{\Lambda_{4}}{3} + \frac{C_{\cal U}}{a^4} .
\ee
For $\gamma = 2/3$ (cosmic strings) the general relativistic term $C_{GR}$ in (\ref{FriedCCC})
scales as cosmic strings in general relativity, while the term with $C_{\lambda}$ scales as
radiation in general relativity (compare \cite{Singh1/3,Parampreet02}), i.e.,
\be
\label{Friedgam23}
\frac{1}{a^2} \left( \frac{da}{dt} \right)^2 =
\frac{C_{\cal U}+C_{\lambda}}{a^4} + \frac{C_{GR}-k}{a^{2}}
+ \frac{\Lambda_{4}}{3}   ,
\ee
with an effective curvature index $k' \equiv k - C_{GR}$.
Then, the problem of writing down exact solutions, which are
elementary, reduces to the repetition of the discussion of Ref.
\cite{Dabrowski96}. We will not be doing this here.
For other values of $\gamma = 4/3; 1; 2$ the terms of the type $1/a^8$
and $1/a^{12}$ appear, and the integration involves hyperelliptic
integrals.

Let us also consider the case of $\gamma = -1/3$ (phantom) \cite{darkenergy}, i.e.,
\be
\label{Friedgam-13}
\frac{1}{a^2} \left( \frac{da}{dt} \right)^2 = C_{GR} a +
C_{\lambda} a^2 - \frac{k}{a^{2}} + \frac{C_{\cal U}}{a^4}
+ \frac{\Lambda_{4}}{3}   .
\ee
A simple general relativistic solution for which
$C_{\lambda} = C_{\cal U} = \Lambda_{4} =
0$, $C_{GR} \equiv C_{ph}$, $k = 0$ (flat models) reads as
\be
1/a(t) = \frac{1}{4} C_{ph} (t - t_0)^2 .
\ee
However, it is more interesting to consider
a spatially closed universe with $k=+1$ which reads as
\be
\label{darkclosed}
a(t) = \sqrt{\frac{-1}{C_{ph}^{\frac{1}{2}} \sin{2(t - t_0)}}}  .
\ee
This immediately shows that the admissible domain for time is:
$\pi/2 \leq (t - t_0) \leq \pi$. It is interesting to notice that
in (\ref{darkclosed}) both initial (Big-Bang) and final (Big-Crunch) singularities
emerge for infinite values of the scale factor (i.e., $\varrho \to \infty$ when
$a(t) \to \infty$). This is because $\varrho = C_{ph} a$ in
a conservation law. The new type of singularities are now commonly called Big-Rip
singularities \cite{darkenergy,phantcos}.

It is easy to verify that the standard FRW limit can be obtained from (\ref{Friedroro})
by taking $1/\lambda \to 0$,
i.e., where the inverse of the brane tension tends to zero. The contribution
from the brane $\rho^2$ term is important only when $\rho > \lambda >
(100 \mathrm{GeV})^{4}$.

The type of cosmology we study (cf. Eq. (\ref{Friedroro})) is a special
case of a generalized brane model which is now commonly called a
Cardassian model \cite{freese02,freese03}. In the Cardassian model
the second term on the right-hand side of the equation (\ref{Friedroro})
has an arbitrary power $n$ of the energy density. For brane models
we study $n = 2$.

In order to study observational tests we now define
dimensionless observational density parameters \cite{AJI+II,AJIII}
\bea
\label{Omegadef}
\Omega_{GR}  &=&  \frac{\kappa_{(4)}^2}{3H^2} \varrho ,
\hspace{15pt}
\Omega_{\lambda}  =  \frac{\kappa_{(4)}^2}{6H^2\lambda} \varrho^2 ,
\hspace{15pt}
\Omega_{\cal U}  =  \frac{2}{\kappa_{(4)}^2 H^2\lambda} {\cal U}
,\nonumber \\
\Omega_{k}  &=&  - \frac{k}{H^2a^2} ,
\hspace{15pt}
\Omega_{\Lambda_{4}}  =  \frac{\Lambda_{(4)}}{3H^2} ,
\eea
where the Hubble parameter $H = \dot{a}/a$, and the deceleration parameter
 $q  =  - \ddot{a}a/\dot{a}^2$ ,
so that the Friedmann equation (\ref{FriedCCC}) can be written down
in the form
\begin{equation}
\label{Om=1}
\Omega_{GR} + \Omega_{\lambda} + \Omega_{k} + \Omega_{\Lambda_{(4)}} + \Omega_{\cal U}
= 1  .
\end{equation}
Note that $\Omega_{\cal U}$ in (\ref{Omegadef}), despite standard radiation term, can either be
positive or negative.
Using (\ref{Omegadef}), the equation (\ref{FriedCCC}) can also be
rewritten as (compare Eq.(10) of \cite{AJI+II})
\begin{equation}
\label{Lambda4}
\Omega_{\Lambda_{4}}  = \frac{3\gamma - 2}{2} \Omega_{GR} +
(3\gamma - 1) \Omega_{\lambda} + \Omega_{\cal U} - q .
\end{equation}
It is also useful to express the curvature of spatial sections
in terms of observational parameters by using (\ref{Om=1}) and (\ref{Lambda4})
\begin{equation}
\label{indexk}
- \Omega_{k} = \frac{3\gamma}{2} \Omega_{GR} +
3\gamma \Omega_{\lambda} + 2\Omega_{\cal U} - q - 1.
\end{equation}
These relations (\ref{Lambda4}) and (\ref{indexk}) may allow to write down observational
quantities  for the brane models to study
their compatibility with astronomical data which
is the main objective of the present paper. Obviously, these quantities depend
on the present densities of the different components
of matter content $\Omega$ given by (\ref{Omegadef}) and their equations of state reflected
by the value of the barotropic index $\gamma$.

\section{Brane models in a two-dimensional phase plane}

For the considerations of this section and without the loss of generality it is useful to
simplify the equation~(\ref{Friedroro}) by assuming  that the parameters
$\kappa_{4}^{2}=1$ and ${}^4 M_{p}^{2} = 8\pi$
which gives
\[
\lambda = \frac{6}{\kappa_{5}^{4}}, \qquad
\Lambda_{4} = \frac{1}{2} \sqrt{\frac{6}{\lambda}}
\Lambda_{5} + \frac{1}{2} \lambda~,
\]
so that the equation~(\ref{Friedroro}) reads as
\begin{equation}
\label{eq:12}
H^{2} = \frac{\rho}{3} \left( 1 + \frac{\rho}{2\lambda} \right)
+ \frac{1}{6} \left( \sqrt{\frac{6}{\lambda}} \Lambda_{5} + \lambda \right)
- \frac{k}{a^{2}} + \frac{2}{\lambda} \frac{{\cal U}_0}{a^{4}} ~,
\end{equation}
and it contains only three independent constant $\lambda$, $\Lambda_{5}$
and ${\cal U}_0$. The last constant comes form the conservation law of dark
radiation (\ref{eq:8}). Without loosing a generality in our subsequent analysis
we also put $\Lambda_{5}=1$ so that the equation~(\ref{eq:12}) takes the form
\begin{equation}
\label{eq:13}
H^2 = \frac{\rho}{3} + \frac{\rho^2}{6\lambda} + \frac{\Lambda_4}{3}
- \frac{k}{a^2} + \frac{C_{\cal U}}{a^4}~.
\end{equation}

As we have mentioned already brane cosmology with the Randall-Sundrum ansatz is mathematically
equivalent to a multifluid Friedmann cosmology. It is governed by the following
dynamical system \cite{Szydlo02}
\begin{eqnarray}
\label{eq:14}
\dot{x} &=& y ~,\\
\dot{y} &=& - \frac{\partial V}{\partial x}
= - \frac{1}{2} \sum_{i} \Omega_{i,0} (3\gamma_i-2)
x^{-(3\gamma_i-1)}~,\nonumber
\end{eqnarray}
where dot denotes differentiation with respect to a new rescaled time variable
$T \colon t \to T \equiv |H_{0}|t$; and $\Omega_{i,0}$ are the density parameters
as defined in (\ref{Omegadef}); $x=a/a_0$, and $p_i = (\gamma_i-1) \rho_i$.
For the sake of this Section we choose that a 4-dimensional brane universe is filled with
{\it non-relativistic} matter (dust) {\it on the brane} only and the cosmological constant, so
that we effectively put $\gamma_{\mathrm{m}}=1$, $\gamma_{\lambda}=2$ (this is due to
the fact that dust contribution which comes from non-relativistic matter scales as stiff-fluid
in standard case - cf. eqs. (\ref{conslaw})-(\ref{FriedCCC}), $\gamma_{\Lambda}=0$.
However, as we already mentioned, the brane ($\varrho^2$) contribution from other types
of fluids will scale differently simulating various types of fluids in standard cosmology.

Here, $V(x)$ plays the role of the potential function for the system
~(\ref{eq:14})
\begin{equation}
\label{eq:15}
V(x) = - \frac{1}{2} \left( \Omega_{k,0} +\sum_{i} \Omega_{i,0}
x^{-(3\gamma_i-2)} \right).
\end{equation}

Let us note that the dynamics can also be represented in terms of a Hamiltonian
dynamical system with the Hamiltonian of the form
\[
\mathcal{H} = \frac{p_{x}^{2}}{2} + V(x)~,
\]
and the system should be considered on the zero-energy level
\begin{equation}
\label{eq:16}
\mathcal{H} = 0.
\end{equation}

\begin{figure}
\includegraphics[width=0.35\textwidth,angle=-90]{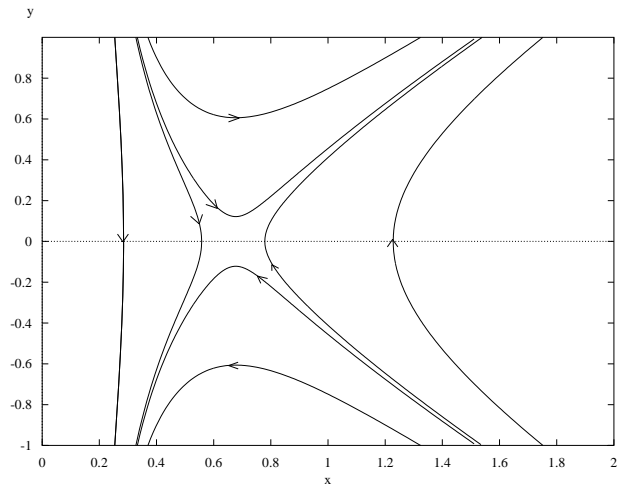}
\caption{The phase portrait of the system~(\ref{eq:14}). There is one critical
point -- a saddle point. This critical point ($x_0,0$) represents
the Einstein Static Universe. The accelerating models lie in the domain $x>x_0$.
Therefore the acceleration region is situated to the right of the saddle point.
The flat model trajectory $\Omega_{k,0}=0$ separates the regions of the models with
negative curvature $\Omega_{k,0}>0$ and positive curvature $\Omega_{k,0}<0$.
The line $x=1$ corresponds to $a=a_0$.}
\label{fig:1}
\end{figure}

The phase portrait of the system~(\ref{eq:14}) for the case of the
dust $\gamma=1$ matter on the brane with
$\Omega_{\Lambda_4,0}=0.69$, $\Omega_{\mathrm{m},0}=0.3$, $\Omega_{\lambda,0}=0.01$
is demonstrated in Fig.~\ref{fig:1}.

\begin{figure}
\includegraphics[width=0.35\textwidth,angle=-90]{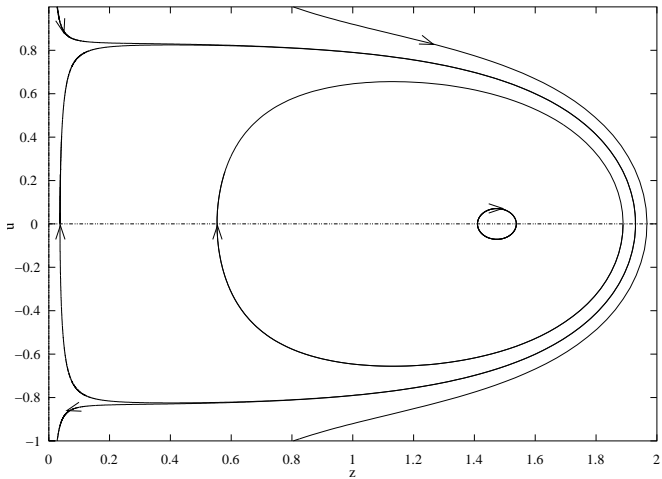}
\caption{The phase portrait of the system~(\ref{eq:14}) in the projective
coordinates $(z,u)$ for the analysis of the behavior of trajectories at
infinity. Two critical points appear at infinity $(0,u_0 \pm)$. They are
of saddle type.}
\label{fig:2}
\end{figure}

The differences in the behavior of trajectories manifest at high densities.
Then, the structure of the dynamical behavior at infinity is modified.
In order to illustrate the behavior of trajectories at infinity,
the system~(\ref{eq:14}) is represented in Fig.~\ref{fig:2} in the projective
coordinates
\begin{eqnarray}
z &=& \frac{1}{x}, \qquad u = \frac{y}{x}, \qquad (z,u) - \mathrm{map}
\label{eq:17a} \\
v &=& \frac{1}{y}, \qquad w = \frac{x}{y}, \qquad (v,w) - \mathrm{map}.
\end{eqnarray}
The two maps $(z,u)\colon z = 0, -\infty < u < \infty$ and $(v,w)\colon v = 0,
-\infty < w < \infty$ cover the behavior of trajectories at the infinity
circles $x = \infty$ and $y = \infty$.

The original system
\begin{eqnarray*}
\dot{x} &=& P(x,y)~, \\
\dot{y} &=& Q(x,y)~,
\end{eqnarray*}
in the projective coordinates $(z,u)$ and after the time reparameterization
$\tau \to \tau_1 \colon d\tau_1 = x d\tau$ takes the form
\begin{eqnarray*}
\dot{z} &=& z P^{*}(z,u)~, \\
\dot{u} &=& Q^{*}(z,u) - uP^{*}(z,u)~,
\end{eqnarray*}
where
\begin{eqnarray*}
P^{*}(z,u) &=& z^{2} P(1/z,u/z) ~,\\
Q^{*}(z,u) &=& z^{2} Q(1/z,u/z) ~,
\end{eqnarray*}
and the dot denotes differentiation with respect to time $\tau_{1}$.

The representation of the dynamics as a one-dimensional Hamiltonian flow allows
to make the classification of possible evolution paths in the configuration
space which is complementary to phase diagrams. It also makes it simpler to
discuss the physical content of the model. Finally, the construction of
the Hamiltonian may allow to study brane quantum cosmology
in full analogy to what is usually done in general relativity.

From equation~(\ref{eq:14}) we can observe that the trajectories are integrable
in quadratures. Namely, from the Hamiltonian constraint $\mathcal{H} = E = 0$
we obtain the integral
\begin{equation}
\label{eq:18}
t-t_{0} = \int_{a_{0}}^{a} \frac{da}{\sqrt{-2V(a)}}.
\end{equation}
Obviously, for a specific form of the potential function~(\ref{eq:15}) we can obtain
exact solutions. Some special cases have been already given in
Section III.

It is possible to make a classification of the evolution paths
by analyzing the characteristic curves which represent the boundary equation
of the domain admissible for motion. For this purpose we consider the
equation of zero velocity, $\dot{a} = 0$ which constitutes the boundary
$\mathcal{M} = \{ a\colon V(a) = 0 \}$.

\begin{figure}
\includegraphics[width=0.35\textwidth,angle=-90]{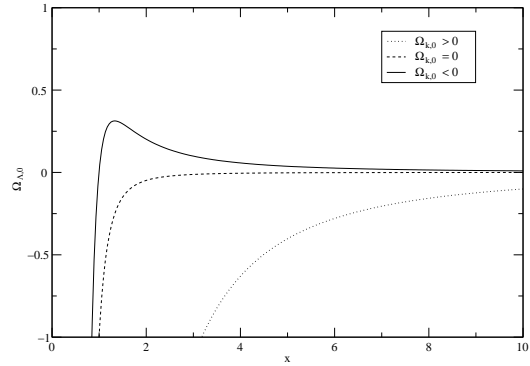}
\caption{The dependence of $\Omega_{\Lambda_4}$ on $x$. The evolutional path are
represented by levels of constant $\Lambda_{4}$. Let us note that the domain
under the characteristic curve $\Omega_{\Lambda_4}(x)$ is non-physical.}
\label{fig:3}
\end{figure}

From equations~(\ref{Lambda4}) and (\ref{indexk}) the cosmological constant
can be expressed as a function of $x$ as follows
\begin{equation}
\label{eq:19}
\Omega_{\Lambda_4,0}(x) = -x^{-2} \left( \Omega_{\mathrm{m},0} x^{-1}
+ \Omega_{\lambda,0} x^{-4} + \Omega_{k,0} \right)~,
\end{equation}
where we have abbreviated $\Omega_{GR}$ by $\Omega_m$.
The plot of $\Omega_{\Lambda_4,0}(x)$ for different $k$ is shown in
Fig.~\ref{fig:3}. Finally, we can consider the evolution path as a level of
$\Omega_{\Lambda_4,0} = \mathrm{const}$ and then we can classify all the
models with  respect to their quantitative properties of the dynamics.

The present experimental estimates based on baryons in clusters,
give $\Omega_{\mathrm{m},0} \sim 0.3$, and the location of
the first peak in the CMB detected by Boomerang
and Maxima suggests a flat universe. This implies that our Universe with the cosmological term
is presently accelerating if (under the assumption of the dust $\gamma=1$
matter on the brane)
\[
\Omega_{\Lambda_4,0} > \frac{1}{3} + \Omega_{\lambda,0}.
\]
The required value of $\Omega_{\lambda,0}$ seems to be unrealistic
(see next section) and therefore the cosmological constant is still
needed to explain the present acceleration of the Universe if we do not accept
phantom type of matter.

\section{A redshift-magnitude relation for brane universes}

Cosmic distance indicators like the luminosity distance, depend sensitively
on the spatial geometry (curvature) and on the present density parameters of
different matter components
and their form of the equation of state. For this reason a redshift-magnitude
relation for distant galaxies is proposed as a potential test for the
Friedmann brane model.

Let us consider an observer located at $r=0$ at the moment $t=t_0$
which receives a light ray emitted at $t=t_1$ from the source of the absolute luminosity $L$
located at the radial distance $r_1$. The redshift z
of the source is related to the scale factor $a(t)$ at the two moments of evolution
by $1+z=a(t_0)/a(t_1) \equiv a_0/a$.
If the apparent luminosity of the source as measured by the observer is $l$,
then the luminosity distance $d_L$ of the source is defined by the
relation
\be
\label{luminosity}
l={L\over 4\pi d_L^2} = {L\over 4\pi} {1 \over (1+z)^2a_0^2r_1^2}.
\ee
The observed and absolute luminosities are
defined in terms of K-corrected apparent and absolute
magnitudes $m$ and ${\cal M}$. When written in terms of $m$ and ${\cal M}$, Eq.(\ref{luminosity}) yields
\be
\label{m(z)}
m(z)={\cal M} + 5\log_{10}
[{\cal D}_L(z)],
\ee
where ${\cal M}=M-5\log_{10}H_0+25$, and ${\cal D}_L \equiv H_0 d_L$
is the dimensionless luminosity distance.
For homogeneous and isotropic Friedmann models one gets \cite{AJIII}
\be
\label{Deelfin}
{\cal D}_L(z) = \frac{\left( 1+z \right)}
{\sqrt{{\cal K}}} S(\chi)
\ee
where $S(\chi)=\sin \chi$ for ${\cal K}=-\Omega_{k,0}$; $S(\chi)=\chi $
for ${\cal K}=0$; $S(\chi)=\sinh \chi$ for ${\cal K}=\Omega_{k,0}$.
From the Friedmann equation (\ref{FriedCCC}) and the form of the FRW metric
we have
\be
\label{chir1}
\chi(z)={1\over a_0 H_0}\int\limits_0^z [f(z^{'})]
^{-1/2}dz^{'},
\ee
where
\bea
f(z) \equiv \left\{\Omega_{\lambda,0}\left(1+z^{'} \right)^{6\gamma} +
\Omega_{GR,0}\left(1+z^{'} \right)^{3\gamma}
\right. \nonumber \\ \left. + \Omega_{k,0}\left(1+z^{'} \right)^2 +
\Omega_{{\cal U},0}\left(1+z^{'}\right)^4 +
\Omega_{\Lambda_{4},0}\right\}.
\eea

In order to understand the variability of $\varrho^2$ contribution
onto the dynamics of the brane universes which is formally
described by the term $\Omega_{\lambda,0}$ let us give a couple of
particular examples. For instance, taking only the non-relativistic
matter on the brane $(\gamma_m = 1)$ we can see that the luminosity
distance can be written down as
\bea
&& d_{L}(z) = \frac{1 + z}{H_0} \times \nonumber \\
&& \int_{0}^{z}
\frac{dz'}{\sqrt{\Omega_{\mathrm{m},0}(1+z')^{3}
+ \Omega_{\lambda m,0}(1+z')^{6} + \Omega_{\Lambda_4,0}}},
\eea
and the `brany' dust contribution $\Omega_{\lambda m,0}$ appears in front of $(1+z')^6$. If
there is only radiation $(\gamma_r=4/3)$ on the brane the
luminostity distance reads as
\bea
&& d_{L}(z) = \frac{1 + z}{H_0} \times \nonumber \\
&& \int_{0}^{z}
\frac{dz'}{\sqrt{\Omega_{\mathrm{r},0}(1+z')^{4}
+ \Omega_{\lambda r,0}(1+z')^{8} + \Omega_{\Lambda_4,0}}}~,
\eea
and the `brany' radiation contribution $\Omega_{\lambda r,0}$ appears in front of $(1+z')^8$.
Another important example is the phantom matter on the
brane $(\gamma=-1/3)$ which gives the luminosity distance in the
form
\be
\label{phandL}
d_{L}(z) = \frac{1 + z}{H_0}
\int_{0}^{z}
\frac{dz'}{\sqrt{\frac{\Omega_{\mathrm{ph},0}}{(1+z')}
+ \frac{\Omega_{\lambda ph,0}}{(1+z')^2} + \Omega_{\Lambda_4,0}}}~,
\ee
and the `brany' phantom contribution $\Omega_{\lambda ph,0}$
appears in front of $(1+z')^{-2}$.

In order to compare various matter on the brane models with the supernova data, we compute the distance
modulus
\[
\mu_{0} = 5\log(d_{L}) + 25
\]
where $d_{L}$ is in Megaparsecs. The goodness of a fit is characterized by
the parameter
\be
\label{chi2}
\chi^{2}=\sum_{i} \frac{{\left(\mu_{0,i}^{0}-\mu_{0,i}^{t}\right)}^2}
{\sigma_{\mu 0,i}^{2} + \sigma_{\mu z,i}^{2}}~.
\ee
In (\ref{chi2}) $\mu_{0,i}^{0}$ is the measured value, $\mu_{0,i}^{t}$
is the value calculated in the model described above, $\sigma_{\mu 0,i}^{2}$
is the measurement error, $\sigma_{\mu z,i}^{2}$ is the dispersion in the
distance modulus due to peculiar velocities of galaxies.

We assume that supernovae measurements come with uncorrelated Gaussian errors
and in this case the likelihood function $\mathcal{L}$ can be determined from
a $\chi^2$ statistic $\mathcal{L} \propto \exp{(-\chi^{2}/2)}$
\cite{Perlmutter99,Riess98}.

\begin{table}
\noindent
\caption{Results of the statistical analysis for the dust matter $(\gamma=1)$ on the brane
for the considered samples of SNIa. First two lines for each sample
are the best fit model and the best fit flat model for the sample. Third line is
the best fit flat model obtained by minimization over
$\mathcal{M}$, while the last line is the best fit flat model with
$\Omega_{m,0}=0.3$ obtained by minimization over $\mathcal{M}$.
}
\begin{tabular}{@{}p{1.5cm}p{0.5cm}rrrrrr}
\hline
Sample & N & $\Omega_{k,0}$ & $\Omega_{\lambda,0}$ & $\Omega_{m,0}$ &
$\Omega_{\Lambda,0}$ & $\chi^2$ &$\mathcal{M}$\\
\hline
  A   &  60 & -0.9 &  0.04  & 0.59 & 1.27 & 94.7&-3.39\\
      &     &  0.0 &  0.091 & 0.00 & 0.90 & 94.7&-3.39\\
      &     &  0.0 &  0.087 & 0.00 & 0.91 & 94.7&-3.40\\
      &     &  0.0 &  0.014 & 0.30 & 0.69 & 95.7&-3.37\\
      &     &      &        &      &      &     &     \\
  C   &  54 & -0.09&  0.092 & 0.00 & 1.00 & 52.5&-3.44\\
      &     &  0.0 &  0.080 & 0.00 & 0.92 & 52.6&-3.44\\
      &     &  0.0 &  0.076 & 0.00 & 0.92 & 52.6&-3.40\\
      &     &  0.0 &  0.004 & 0.30 & 0.70 & 53.3&-3.37\\
      &     &      &        &      &      &     &     \\
  K6  &  58 & -1.26& -0.020 & 0.92 & 1.36 & 55.0&-3.52\\
      &     &  0.0 &  0.076 & 0.00 & 0.92 & 55.2&-3.52\\
      &     &  0.0 &  0.072 & 0.00 & 0.93 & 55.2&-3.53\\
      &     &  0.0 & -0.004 & 0.30 & 0.70 & 55.8&-3.51\\
      &     &      &        &      &      &     &     \\
  K3  &  54 & -0.74& -0.015 & 0.62 & 1.14 & 60.3&-3.48\\
      &     &  0.0 &  0.056 & 0.02 & 0.92 & 60.4&-3.48\\
      &     &  0.0 &  0.059 & 0.10 & 0.93 & 60.4&-3.48\\
      &     &  0.0 & -0.011 & 0.30 & 0.71 & 61.0&-3.46\\
      &     &      &        &      &      &     &     \\
  TB1 & 218 & -0.02&  0.087 & 0.00 & 0.93 &203.9&15.925\\
      &     &  0.0 &  0.085 & 0.00 & 0.92 &203.9&15.925\\
      &     &  0.0 &  0.078 & 0.00 & 0.92 &203.5&15.905\\
      &     &  0.0 &  0.011 & 0.30 & 0.69 &208.3&15.925\\
      &     &      &        &      &      &     &     \\
  TB2 & 120 &  0.82& -0.010 & 0.00 & 0.19 & 89.5&15.925\\
      &     &  0.0 & -0.020 & 0.54 & 0.48 & 89.6&15.925\\
      &     &  0.0 & -0.013 & 0.33 & 0.68 & 90.3&15.905\\
      &     &  0.0 & -0.002 & 0.30 & 0.70 & 90.3&15.905\\
      &     &      &        &      &      &     &     \\
  TB3 &  98 & -0.18&  0.104 & 0.00 & 1.08 &112.1&15.925\\
      &     &  0.0 &  0.084 & 0.00 & 0.92 &112.5&15.925\\
      &     &  0.0 &  0.088 & 0.13 & 0.78 &111.8&16.035\\
      &     &  0.0 &  0.044 & 0.30 & 0.66 &112.4&16.035\\
\hline
\end{tabular}
\end{table}

\section{Brane universes tested by supernovae}

We first test brane models using the sample A of Perlmutter SN Ia data.
In order to avoid any possible selection effects we work with the full
sample without excluding any supernova from that sample. It means that our
basic sample is Perlmutter sample A. We checked our analysis using Permutter's
samples B and C but not significance difference was obtained.
We estimate the model parameters using the best fit procedure. In the
statistical analysis we also use the maximum likelihood method \cite{Riess98}.

Firstly, we will study the case $\gamma = 1$ (dust on the brane;
we will label $\Omega_{GR}$ by $\Omega_m $).
The case $\gamma = 2/3$
(cosmic strings on the brane) has recently been studied in \cite{Singh1/3} where, in fact,
$\Omega_{\cal U}$ and $\Omega_{\lambda}$ were neglected and
where the term $\Omega_{m,0} (1+z^{'})^3$ was
introduced in order to admit dust matter on the brane. This case was already presented in
a different framework in Ref. \cite{AJIII}.
Secondly, we will study the case $\gamma = -1/3$ (phantom on the brane \cite{darkenergy}
- we will label this type of matter with $\Omega_{ph}$ instead of $\Omega_{GR}$).

Let us first estimate the value of $\mathcal{M}$ from the full
sample of 60 supernovae. For the flat $\Lambda$CDM model we obtained
$\mathcal{M} = -3.39$. We use Perlmutter data and method of maximum likelihood
estimation on this data to estimate the different cosmological parameters of
interest, i.e.\ pairs ($\Omega_{\mathrm{m},0}$, $\Omega_{\lambda,0}$),
($\Omega_{\mathrm{m},0}$, $\Omega_{\Lambda,0}$).

\begin{figure}
\includegraphics[width=0.42\textwidth]{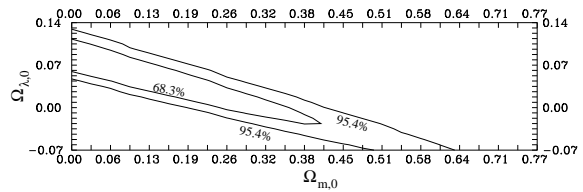}
\caption{Confidence levels in the plane
$(\Omega_{\mathrm{m},0} , \Omega_{\lambda,0})$ minimized over $\mathcal{M}$
for the flat model, and with
$\Omega_{\Lambda_4,0}=1-\Omega_{\mathrm{m},0}-\Omega_{k,0}-\Omega_{\lambda,0}$.
The figure shows the ellipses of the preferred values of
$\Omega_{\mathrm{m},0}$ and $\Omega_{\Lambda_4,0}$. The results prefer
positive values of $\Omega_{\lambda,0}$, but negative values are
allowed as well (Perlmutter sample A).}
\label{fig:4}
\end{figure}

The result of statistical analysis is presented in some figures and in Table I.
Fig.~\ref{fig:4} illustrates the confidence level as a function of
($\Omega_{\mathrm{m},0}$, $\Omega_{\lambda,0}$) for the flat model
($\Omega_{k,0}=0$) minimized over $\mathcal{M}$ with
$\Omega_{\Lambda,0}=1-\Omega_{\mathrm{m},0}-\Omega_{k,0}-\Omega_{\lambda,0}$.
In present cases we formally assume that both positive and negative values of
$\Omega_{\lambda,0}$ are mathematically possible although the negative values are in fact
only possible if we admit the timelike extra dimensions \cite{Shtanov}.
We show that the preferred intervals for $\Omega_{m,0}$ and
$\Omega_{\lambda,0}$ are $\Omega_{m,0}<0.4$ and
$\Omega_{\lambda,0}>0$.

We repeat our analysis for the case of $\Omega_{k,0} \ne 0$.
The results of the statistical analysis are presented in Table I.
Two upper lines for each sample are best fit model and best fit flat
model for sample with fixed $\mathcal{M}$ as obtained for the flat $\Lambda$CDM.
Third line is best fit flat model obtain with  minimization over $\mathcal{M}$
while last line is best fit flat model with  $\Omega_{m,0}=0.3$ (preffered by
present extragalactic data \cite{Peebles03}) obtained with  minimization
over $\mathcal{M}$.

One should note that the
result obtained for Perlmutter's sample C is a little bit different than
the one
obtained in preliminary analysis \cite{Godlowski04} for $\mathcal{M}=-3.39$
($\Omega_{k,0}=0.$, $(\Omega_{\mathrm{m},0}=0.21$,
$\Omega_{\lambda,0}=0.048$ $\Omega_{\Lambda,0}=0.75$). It reflects the sensitivity
of the results to changes of $\mathcal{M}$ parameter.

The Perlmutter data were gathered four years ago, hence it would be
interesting to use more recent supernovae data as well.
We decided to test our  model using this new sample of supernovae.
Recently Knop et al. \cite{Knop03} have reexamined the Permutter's data with host-galaxy
extinction correctly assessed.
The mentioned authors distinguished  few subsets of supernovae from this sample.
We consider two of them. The first is a subset of 58 supernovae with extinction
correction (Knop subsample 6; hereafter K6) and the second one a sample of 54 supernovae
with low extinction (Knop subsample 3; hereafter K3). Sample C and K3 are
similarly constructed because both contain only low extinction supernovae.

Another sample was presented by Tonry {\it et al.} \cite{Tonry03} who
collected a large number of supernovae published by different authors
and added eight new high redshift SN Ia. This sample of 230 Sne Ia
was recalibrated with consistent zero point. Whenever it was
possible, the extinctions estimates and distance fitting were recomputed. However, none
of the methods was able to be applied to all supernovae (for details see
Table~8 in \cite{Tonry03}). This sample was improved by Barris
who added 23 high redshift supernovae including 15 at $z \ge 0.7$ doubling the
published number of object at this redshifts \cite{Barris03}.

Despite of these problems, the analysis of our model using
this sample seems to be interesting. We decided to analyze
the sample of 218 SNe Ia (hereafter sample TB1) which consists of low extinction
supernovae only (median $V$ band  extinction $A_V<0.5$).

Tonry and Barris \cite {Tonry03,Barris03} presented redshift and luminosity
distance observations for their sample of supernovae. Therefore,
Eqs. (23a) and (23b) should be modified \cite{Williams03}:
\begin{equation}
\label{eq:13a}
m-M = 5\log_{10}(\mathcal{D}_L)_{\mathrm{Tonry}}-5\log_{10}65 + 25
\end{equation}
and
\begin{equation}
\label{eq:13b}
\mathcal{M}=-5\log_{10}H_0+25.
\end{equation}
For $H_0=65$ km s$^{-1}$ Mpc$^{-1}$, we obtain $\mathcal{M}=15.935$.

The results obtained for the flat model with the Tonry/Barris sample of
218 SNIa is very similar to that obtained for flat model with the Perlmutter
sample. However, even then we allowed $\Omega_{k,0} \ne 0$ than for TBI sample
we obtained that preffered model of the universe is nearly flat one, which
is in agreement with CMBR data. It is the advantage of our model in comparision
to $\Lambda$CDM model, where \cite{Riess98,Perlmutter99} the high
negative value of $\Omega_{k,0}$ was the best fit, although
$\Omega_{k,0}$ is also statistically admissible. In order to find the curvature, they
additionally used the data from CMBR and extragalactic astronomy.

We also confront our model against Knop's sample. While for the
flat model we obtained similar results than for the previous samples,
in the case without of any priors on $\Omega_{k,0}$ Knop's sample
prefers its highly negative values together with the negative values of $\Omega_{\lambda,0}$.
Note, however that it is an unphysical case, because the brane tension should be positive
unless timelike extra dimensions are admitted \cite{Shtanov}. One should note that with this
assumption the Knop's sample
suggests $(\Omega_{\mathrm{m},0}<0.3$ which is in agreement with
the result obtained for this sample in the case of $\Lambda$CDM model \cite{Knop03}.

\begin{figure}
\includegraphics[width=0.42\textwidth]{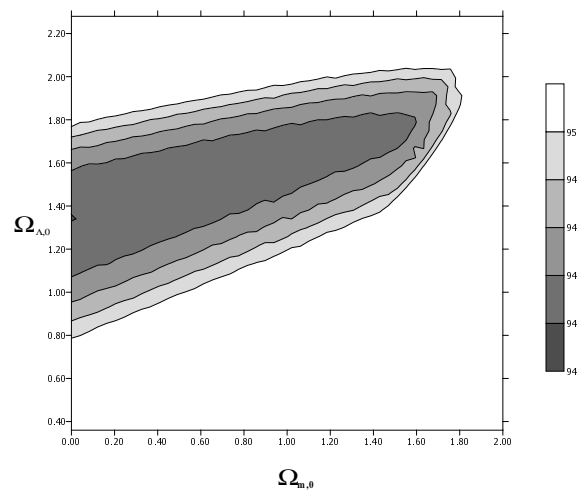}
\caption{The levels of constant $\chi^{2}$ in the plane
$(\Omega_{\mathrm{m},0},\Omega_{\Lambda_4,0})$ marginalized over $\mathcal{M}$,
and with
$\Omega_{\Lambda_4,0}=1-\Omega_{\mathrm{m},0}-\Omega_{k,0}-\Omega_{\lambda,0}$
The figure shows the preferred values of
$\Omega_{\mathrm{m},0}$ and $\Omega_{\Lambda_4,0}$ (Perlmutter sample A).}
\label{fig:5}
\end{figure}

Applying the marginalization procedure over $\Omega_{k,0}$,
$\Omega_{\lambda,0}$, $\mathcal{M} $ for the Perlmutter sample A we find
the lowest value of $\chi^2$ for each pair of values
$(\Omega_{\mathrm{m},0},\Omega_{\Lambda,0})$ as shown in Fig.~\ref{fig:5}.
The favoured intervals for   $(\Omega_{\mathrm{m},0},
\Omega_{\Lambda,0})$ are $\Omega_{m,0}<0.5$ and $\Omega_{\Lambda,0}\simeq 1.3$.
As the best fit we obtain: $\Omega_{k,0}=-0.48$, $\Omega_{\mathrm{m},0}=0$,
$\Omega_{\lambda,0}=0.14$,  and $\mathcal{M}=-3.43$.
There is a marginal difference between that result and the one obtained for
$\mathcal{M}=-3.39$. Another example is that for $\mathcal{M}=-3.39$ and $\Omega_{k,0}=-0.20$,
$(\Omega_{\mathrm{m},0}=0$, $\Omega_{\lambda,0}=0.118$ $\Omega_{\Lambda,0}=1.08$
and for the case noted in the first line of tab I we obtain nearly the same
$\chi^{2}$ value.

In fact, we obtained a 3D ellipsoid in a 3d parameter space $\Omega_{m,0}$,
$\Omega_{\lambda,0}$, $\Omega_{\Lambda_{4},0}$. Then, we have more freedom than in
the case of analysis of Ref. \cite{Perlmutter99}, where there was only an ellipse
in a 2D parameter space $\Omega_{m,0}$ and $\Omega_{\Lambda_{4},0}$.
It clearly demonstrates that the results obtained from the
best-fit analysis should be supported by the analysis of the confidence levels
for the parameter intervals obtained from maximum likelihood method.
Support from CMBR and extragalactic astronomy results may also be useful.

\begin{figure}
\includegraphics[width=0.42\textwidth]{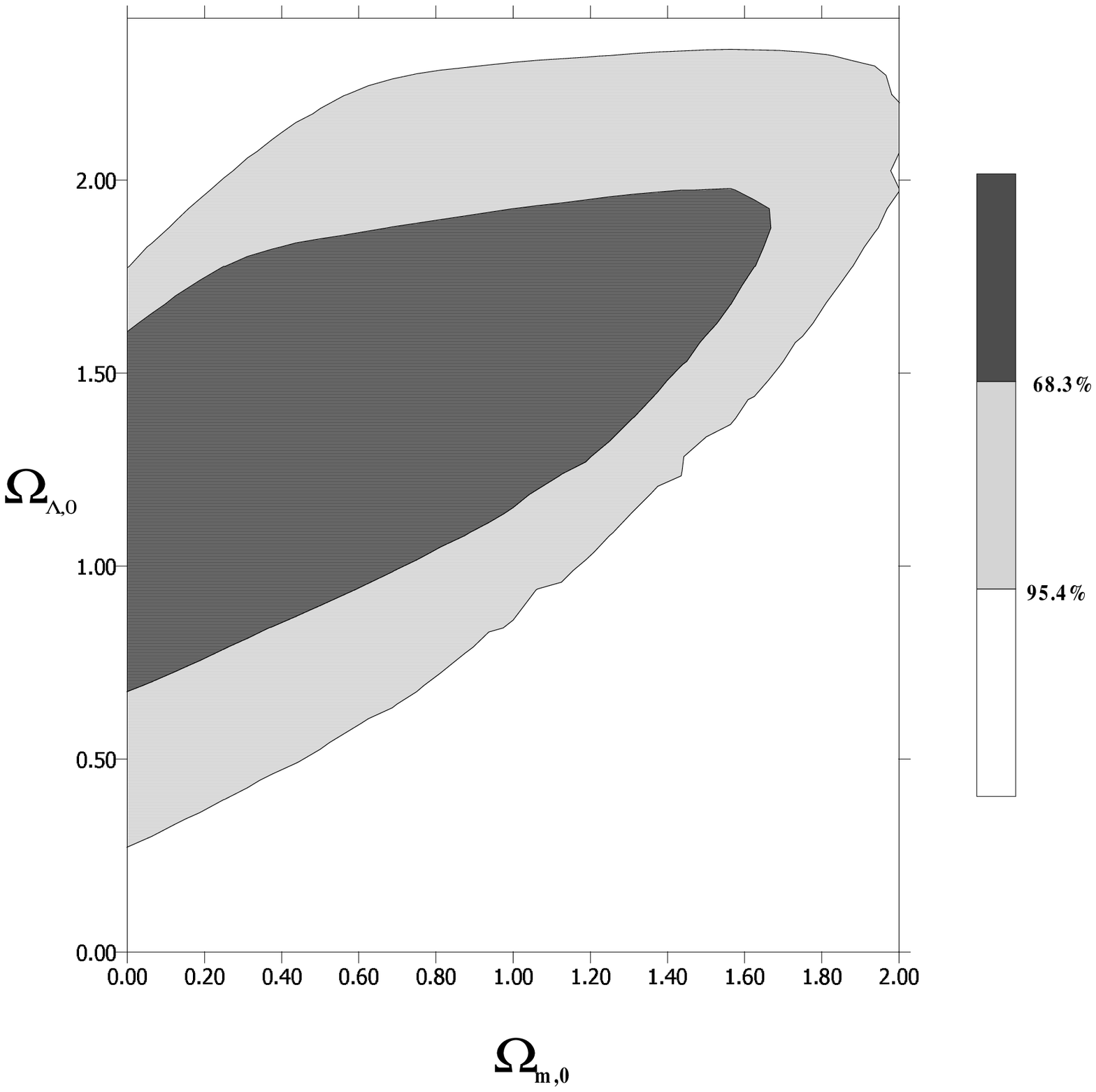}
\caption{Confidence levels in the plane
$(\Omega_{\mathrm{m},0} , \Omega_{\Lambda_4,0})$ minimized over $\mathcal{M}$,
and with
$\Omega_{\Lambda_4,0}=1-\Omega_{\mathrm{m},0}-\Omega_{k,0}-\Omega_{\lambda,0}$.
The figure shows the ellipses of the preferred values of
$\Omega_{\mathrm{m},0}$ and $\Omega_{\Lambda_4,0}$ (Perlmutter sample A).}
\label{fig:6}
\end{figure}

\begin{figure}
\includegraphics[width=0.42\textwidth]{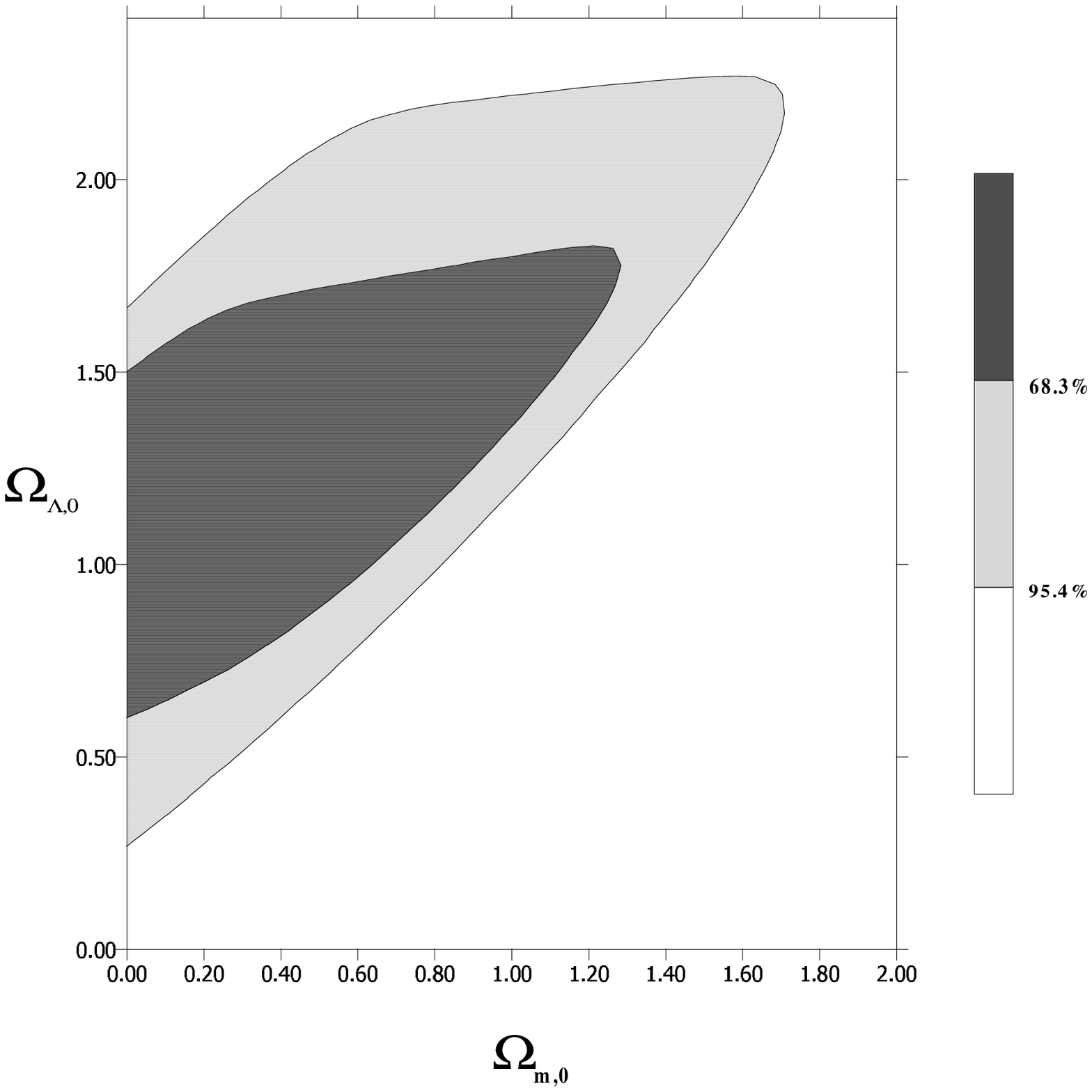}
\caption{Confidence levels in the plane
$(\Omega_{\mathrm{m},0} , \Omega_{\Lambda_4,0})$, $\mathcal{M}=-3.39$,
and with
$\Omega_{\Lambda_4,0}=1-\Omega_{\mathrm{m},0}-\Omega_{k,0}-\Omega_{\lambda,0}$.
The figure shows the ellipses of the preferred values of
$\Omega_{\mathrm{m},0}$ and $\Omega_{\Lambda_4,0}$ (Perlmutter sample A).}
\label{fig:6o}
\end{figure}

In the case of $\Omega_{k,0} \ne 0$, the confidence level for values of
pairs $(\Omega_{\mathrm{m},0} ,\Omega_{\Lambda,0})$ (Perlmutter sample A)
are shown in the standard way after minimizing over $\mathcal{M}$,
$\Omega_{k,0}$, $\Omega_{\lambda,0}$ in
Fig.~\ref{fig:6}. One can see that a non-zero cosmological constant
is required in our model.
This analisis was repeated with fixed value of $\mathcal{M}=-3.39$
(Fig.~\ref{fig:6o}). These figures shows that procedure of the
minimizing over $\mathcal{M}$, however important, is not crucial for
selecting preferred intervals of $(\Omega_{\mathrm{m},0},
\Omega_{\Lambda,0})$ in our model.

\begin{figure}
\includegraphics[width=0.42\textwidth]{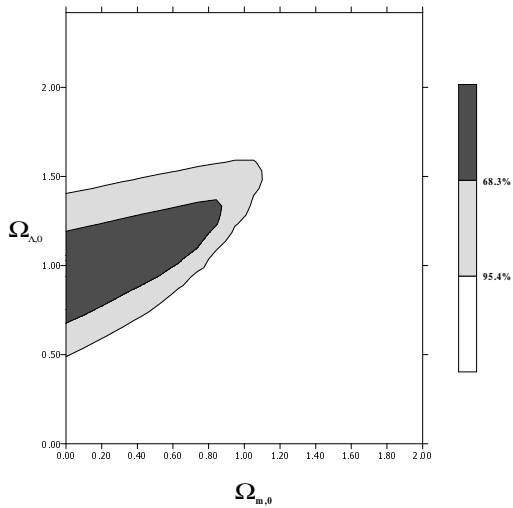}
\caption{Confidence levels in the plane
$(\Omega_{\mathrm{m},0} , \Omega_{\Lambda_4,0})$ for the sample Tonry/Barris
218 low extincted SNIA.}
\label{fig:6a}
\end{figure}

In Fig.~\ref{fig:6a} we presented confidence levels in the plane
$(\Omega_{\mathrm{m},0},\Omega_{\Lambda,0})$ for TBI sample.
The allowed area is in agreement with that obtained for the Perlmutter
sample, but significantly smaller. The earlier conclusion that the non-zero
cosmological constant is required is confirmed. However, as we
will see further, the admission of strongly negative pressure matter
(phantom) on the brane does not require cosmological constant.

\begin{figure}
\includegraphics[width=0.42\textwidth]{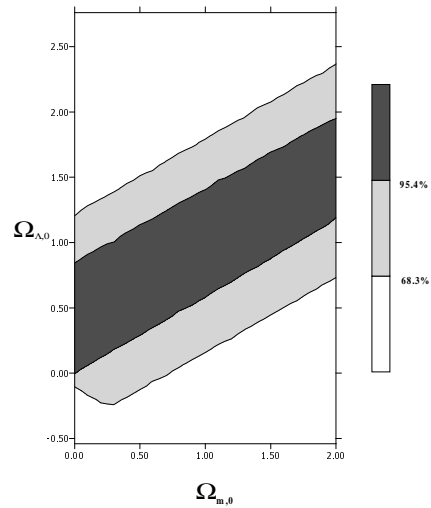}
\caption{Confidence levels in the plane
$(\Omega_{\mathrm{m},0} , \Omega_{\Lambda_4,0})$ for the sample Tonry/Barris
120 low extincted SNIA with $z \le 0.25$. The figure illustrate that with
sample of low redshift supernovae we obtain that model
without cosmological constant is statistically admissible.}
\label{fig:6b}
\end{figure}

\begin{figure}
\includegraphics[width=0.42\textwidth]{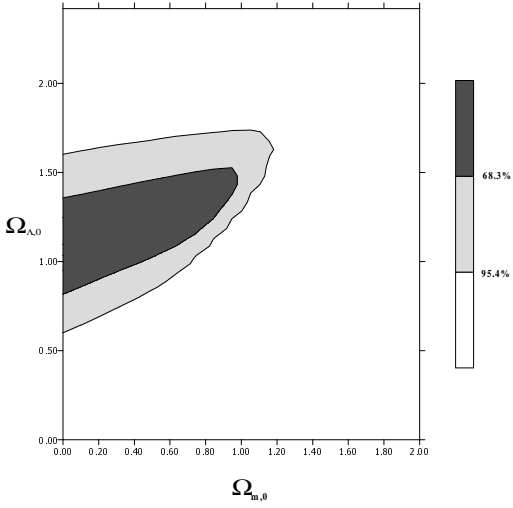}
\caption{Confidence levels in the plane
$(\Omega_{\mathrm{m},0} , \Omega_{\Lambda_4,0})$ for the sample Tonry/Barris
of 98 low extincted SNIa with $z>0.25$.}
\label{fig:6c}
\end{figure}

Padmanabhan and Choundhury \cite{Padmanabhan02,Choundhury03}
suggest that dividing the sample finto the low and high redshift supernovae
gives interesting results.
They obtained that although the full data sets of SNIa strongly
rules out models without dark energy, the high and low redshift data sets individually,
admit decelerating models with zero dark energy \cite{Padmanabhan02,Choundhury03}.
We decide to check this result it the case of our model. We divide TB1 sample
for 120 low redshift SNIa with $z \le.0.25$ (TB2 sample) and 98 high redshift
SNIa with $z >0.25$ (sample TB3). The results are presented in Table I and
in Figs.~\ref{fig:6b} - Fig.~\ref{fig:6c}. One can see that the preferred values
of the model parameters are different in both cases. For low redshifts
$z \le 0.25$ the data sets taken individually results in
decelerating models without cosmological constant, while for high redshift
$z \le 0.25$ data sets it requires the non-zero cosmological constant.
It confirms that pure low redshift data cannot be used for discrimination
between cosmological models effectively (see \cite{Padmanabhan02}).

\begin{figure}
\includegraphics[width=0.38\textwidth]{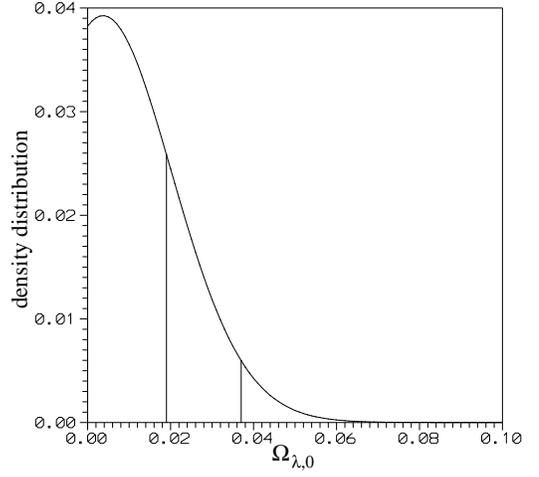}
\caption{The density distribution for $\Omega_{\lambda,0}$ in the model
with (only positive) brane fluid present. We obtain the limit
$\Omega_{\lambda,0} < 0.019$ at the confidence level $68.3$, and
$\Omega_{\lambda,0} < 0.037$ at the confidence level $95.4$
($\Omega_{\mathrm{m},0}=0.3$, Perlmutter sample A)}
\label{fig:7}
\end{figure}

On should note, that the present extragalactic data for galaxy clusters
(cluster baryon fraction) with CMB anisotropy measurement prefer a flat model
($\Omega_{k,0} $) with $\Omega_{m,0} \simeq 0.3$ \cite{Peebles03}.
For a deeper statistical analysis of brane expansion scenario
in explaining the currently accelerating universe we consider 1D plot
of the density distribution of $\Omega_{\lambda,0}$. From this
analysis one can obtain the limits at the $1 \sigma$ level.
Fig.~\ref{fig:7} shows the density distribution for $\Omega_{\lambda,0}$
in the flat model with $\Omega_{\mathrm{m},0} = 0.3$. This distribution is
obtained from the marginalization over $\mathcal{M}$.
On the base of the noted above results we assumed that "true" value
of $\mathcal{M}$ is in the interval $[-3.37,-3.44]$ and that interval
we take into account during the marginalization procedure.
One can conclude that with the probability of $68.3$ we get
$\Omega_{\lambda,0} < 0.019$ while with the probability $95.4$ we get
$\Omega_{\lambda,0} < 0.037$.

\begin{figure}
\includegraphics[width=0.38\textwidth]{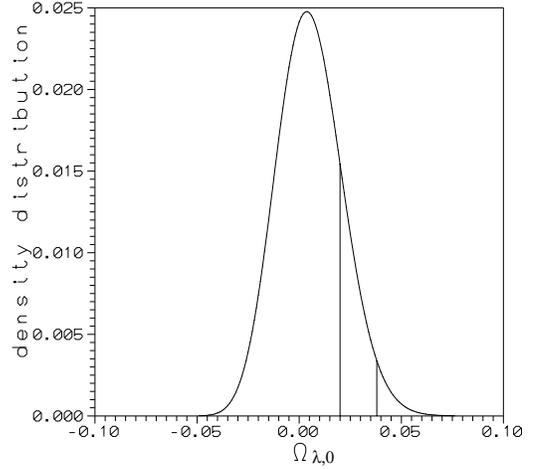}
\caption{The density distribution for $\Omega_{\lambda,0}$ in the brane
model. We obtain that $\Omega_{\lambda,0} < 0.020$ at $1 \sigma$ level
and $\Omega_{\lambda,0} < 0.038$ at $2 \sigma$ level. Both positive and
negative values of $\Omega_{\lambda,0}$ are formally possible.
For simplicity, we only mark limits on the positive side.
($\Omega_{\mathrm{m},0}=0.3$, Perlmutter sample A).}
\label{fig:8}
\end{figure}

If we formally consider the possibility of a negative value of
$\Omega_{\lambda,0}$  which can be interpreted
as the negative brane tension with dust on the brane
\cite{Szydlo02} the corresponding distribution function is shown in
Fig.~\ref{fig:8}. One can see that $\Omega_{\lambda,0} < 0.020$ on
the $1 \sigma$ confidence level and $\Omega_{\lambda,0} < 0.038$ on the
$2 \sigma$ confidence level with preferred value $\Omega_{\lambda,0}=0.004$.
However, the error in estimation is so large that possibility that
$\Omega_{\lambda,0}=0$ cannot be excluded.
One should note, that when we formally increase the analyzed interval of
$\mathcal{M}$ to $[-3.29,-3.49]$ we obtain the preferred value
$\Omega_{\lambda,0}=0.013$. Also, the errors in estimation increase and
$\Omega_{\lambda,0} < 0.037$ on the $1 \sigma$ confidence level
and $\Omega_{\lambda,0} < 0.061$ on the $2 \sigma$ confidence level.
It demonstrates  the sensitivity of the results of estimation of
$\Omega_{\lambda,0}$ with respect to small changes of $\mathcal{M}$ parameter.

In the near future the
SNAP mission is expected to observe about 2000 type Ia supernovae each year,
over a period of three years \cite{snap}. Therefore it could be possible to verify hypothesis
that $\Omega_{\lambda,0}>0$ because errors in the estimation of
$\Omega_{\lambda,0}$ will decrease significantly.
We test how large number of new data
should influence on errors in estimation of $\Omega_{\lambda,0}$.
We assume that the Universe is flat with $\Omega_{\mathrm{m},0}=0.28$,
$\Omega_{\lambda,0}=0.01$ and ${\cal M}=-3.39$.
For the model with dust matter on the brane we generate the sample of 1000 supernovae
randomly distributed in the redshift range $0.01 <z<2$.
We assume a Gaussian distribution of uncertainties in the measured values of m and z.
The errors in redshifts $z$are of order $1\sigma=0.002$ while the uncertainty
in a measurement of the magnitude $m$ is assumed as $1\sigma=0.15$. The systematic
uncertainty limits is $\sigma_{sys}=0.02$ mag at $z=1.5$ \cite{Alam03}
which means that $\sigma_{sys}(z)=(0.02/1.5)z$.
For the sample generated in such a way we shoul now repeat our analysis.
The error for $\Omega_{\lambda,0} \simeq 0.0007$ on the confidence level of $68.3$,
while $\Omega_{\lambda,0} \simeq 0.0013$ is on the confidence level $95.4$.
It is clearly confirmed that the error in measurement of $\Omega_{\lambda,0}$
from supernovae data will decrease significantly in the new future.

In Fig.~\ref{fig:9} we present the plot of residuals of redshift-magnitude
relationship for the supernovae data for $\Omega_{\mathrm{m},0}=0.3$.
We present the residuals between the Einstein-de Sitter model and Perlmutter
model, best fitted flat brane model and best fitted brane model.
As a result, with the increasing impact of $\Omega_{\lambda,0}$
(higher $\Omega_{\lambda,0}$) the high-redshift supernovae should be
brighter than the expected by the Perlmutter model.
Let us note that for the best-fit
value of $\Omega_{\lambda,0} = 0.020$, the difference between the brane model
and the Perlmutter model should be detectable for $z > 1.2$ (because
$ \Delta m \geq 0.2$). One should observe that there is a small difference
between the best fit flat model (with minimal value of $\chi^2$)
where we obtain $\Omega_{\lambda,0}=0.006$ and the value obtained from
the minimization procedure $\Omega_{\lambda,0}=0.004$ (see Fig.~\ref{fig:8}).

For comparison we presented the best fits (Fig.~\ref{fig:10}) without a
specifically assumed value of $\Omega_{m,0}$. This latter case is,
characterized by greater expected differences between the brane model and
the Perlmutter model for high redshift supernovae.

The above prediction for brane models that the high redshift supernovae are
{\it brighter than expected} in $\Lambda$CDM (Perlmutter) model could also
be detectable by the new SNAP data. It gives a possibility to discriminate between
predictions of the two models. Of course similar effect is expected for galaxies,
but for an extended object such an effect is much more difficult to detect
than for the point sources.

\begin{figure}
\includegraphics[width=0.38\textwidth]{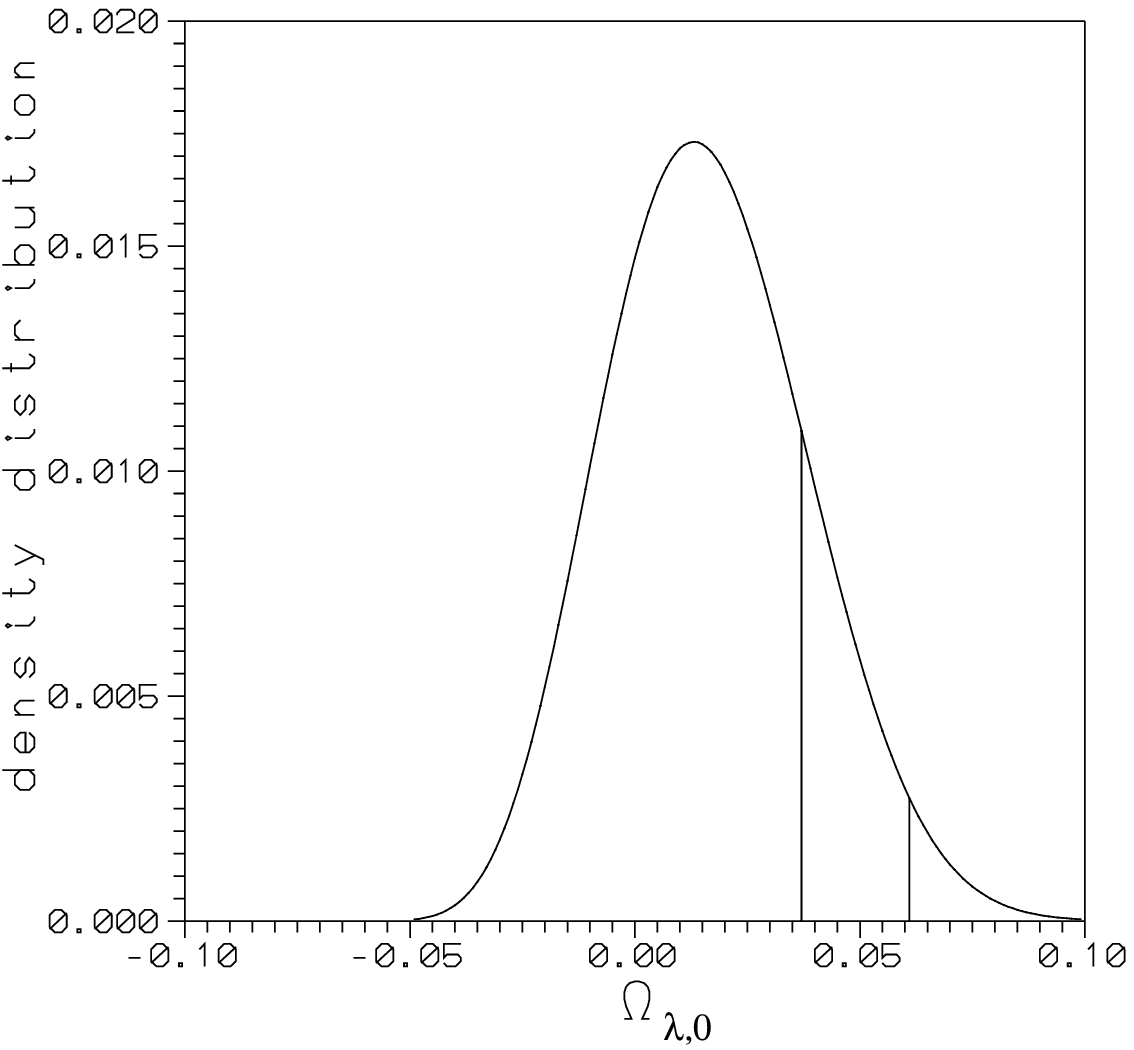}
\caption{The density distribution for $\Omega_{\lambda,0}$ in the brane
model (with ${\cal M}$ in the interval $[-3.29; 3.49] $). We obtain that
$\Omega_{\lambda,0}=0.013$ and
$\Omega_{\lambda,0} < 0.037$ at $1 \sigma$ level
and $\Omega_{\lambda,0} < 0.061$ at $2 \sigma$ level. Both positive and
negative values of $\Omega_{\lambda,0}$ are formally possible.
For simplicity, we only mark limits on the positive side.
($\Omega_{\mathrm{m},0}=0.3$, Perlmutter sample A).}
\label{fig:8a}
\end{figure}

\begin{figure}
\includegraphics[width=0.38\textwidth]{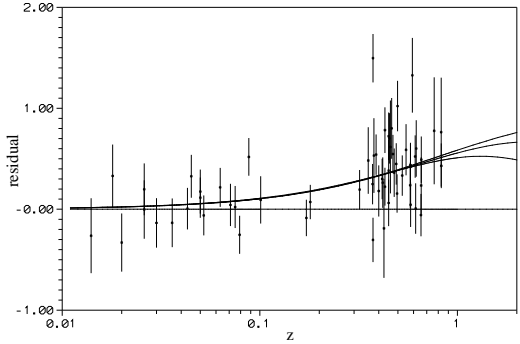}
\caption{Residuals (in mag) between the Einstein-de Sitter model and:
the Einstein-de Sitter model itself (zero line), the Perlmutter flat model
(upper curve), the $\Omega_{\mathrm{m},0}=0.3$ the best-fit flat brane
model ($\Omega_{\lambda,0} = 0.006$ and $\Omega_{k,0} = 0$)
(upper-middle curve), and $\Omega_{\mathrm{m},0}=0.3$
best-fit brane model ($\Omega_{\lambda,0} = 0.020$,
$\Omega_{k,0}=-0.1$ (lower-middle curve).}
\label{fig:9}
\end{figure}

\begin{figure}
\includegraphics[width=0.38\textwidth]{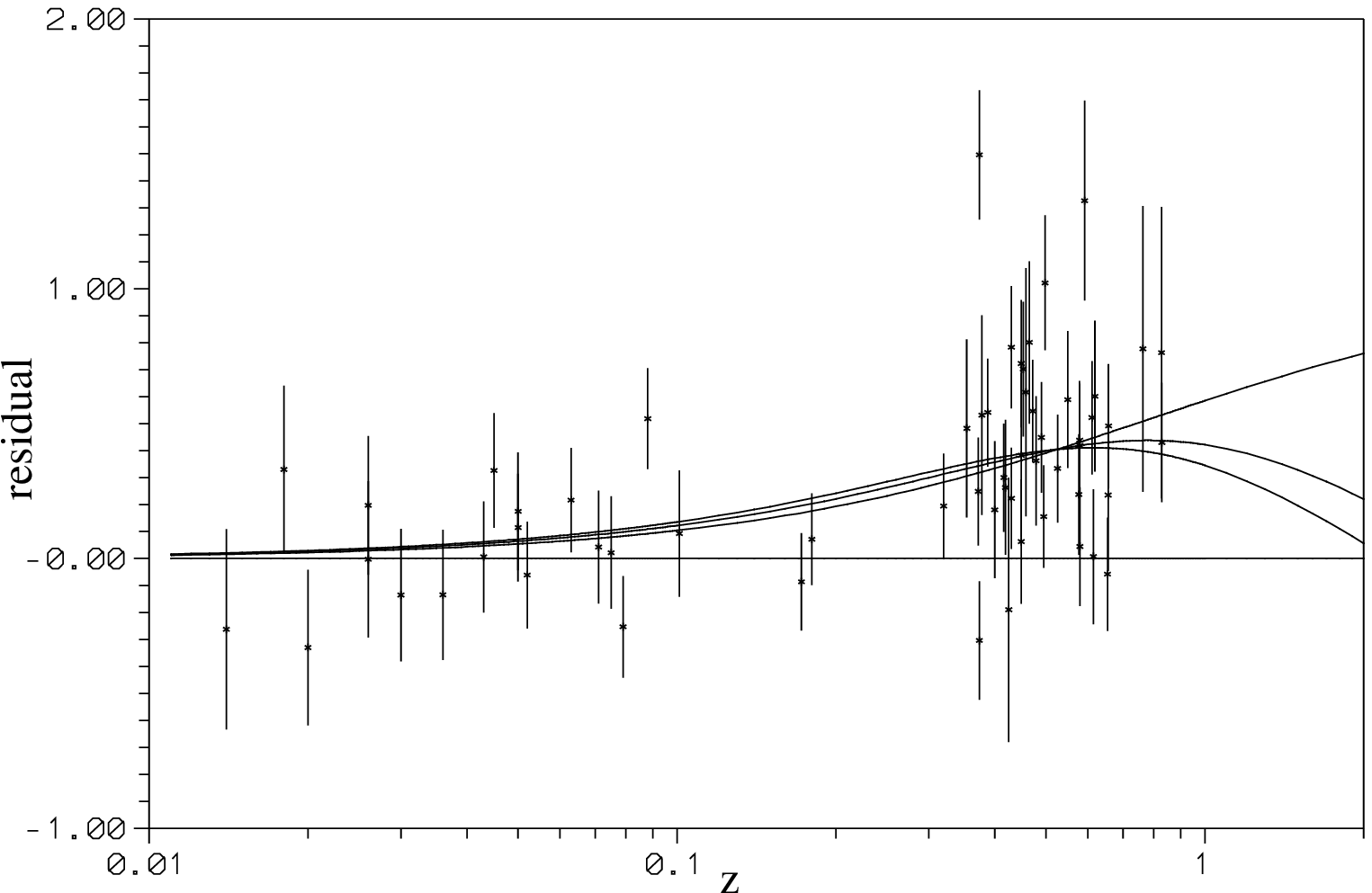}
\caption{Residuals (in mag) between the Einstein-de Sitter model and:
the Einstein-de Sitter itself (zero line), the Perlmutter flat model
(upper curve), the best-fit flat brane model (upper-middle curve),
$\Omega_{k,0}=0$, $\Omega_{\mathrm{m},0}=0.25$, $\Omega_{\lambda,0}=0.02$,
$\Omega_{\Lambda_4,0}=0.73$ and the best-fit brane (lower-middle curve)
$\Omega_{k,0}=-0.9$, $\Omega_{\mathrm{m},0}=0.59$, $\Omega_{\lambda,0}=0.04$
$\Omega_{\Lambda_4,0}=1.27$.}
\label{fig:10}
\end{figure}

\begin{figure}[h]
\includegraphics[angle=0,scale=.46]{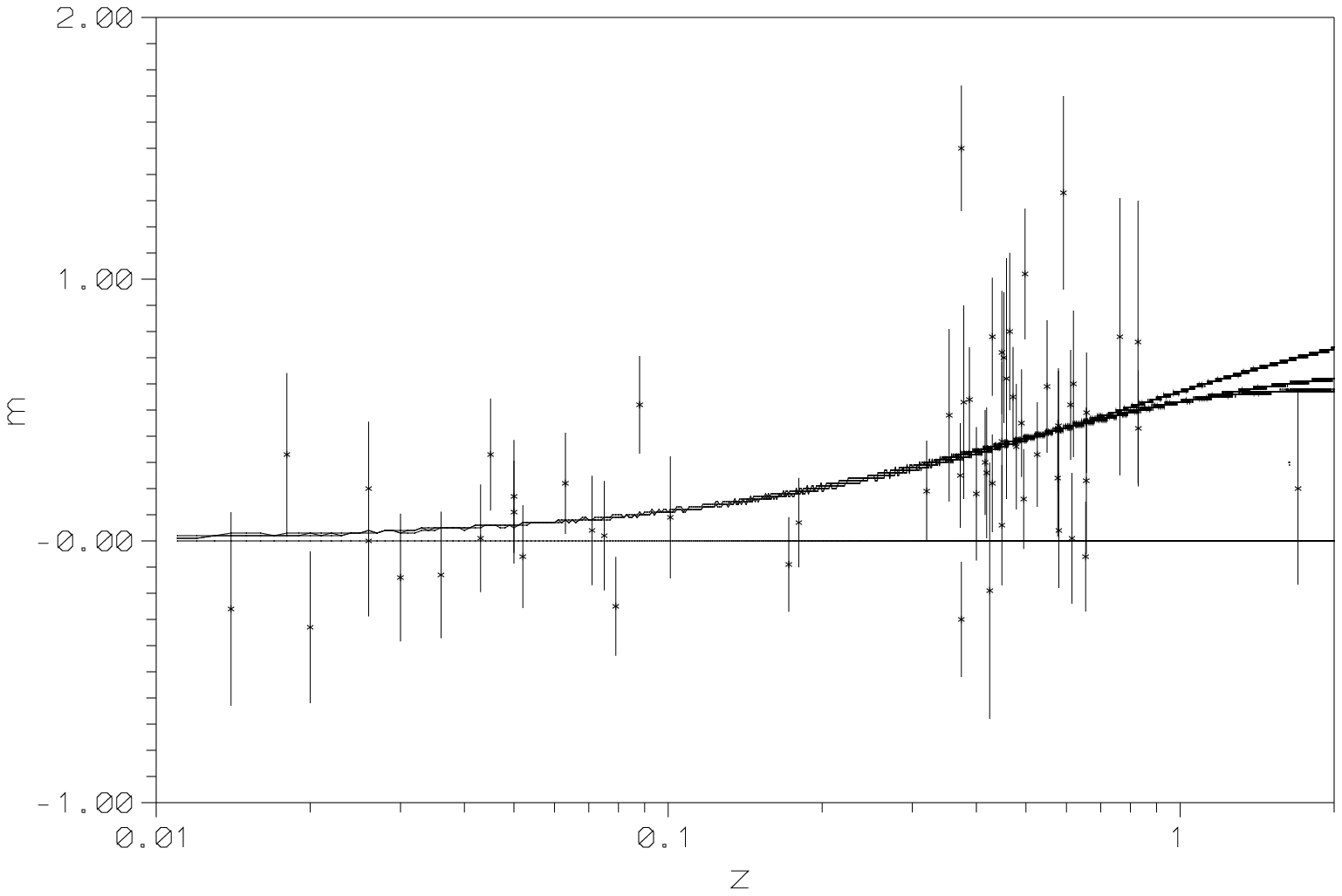}
\caption{Residuals (in mag) in redshift-magnitude relation for $\gamma = -1/3$ brane universes
(phantom matter on the brane). The top line is the best-fit flat model of Ref. \cite{Perlmutter99} with $\Omega_{m,0}=0.28$,
$\Omega_{\Lambda_{(4)},0}=0.72$. The bottom line is a pure flat model with
$\Omega_{\Lambda_{(4)},0}=0.$ Between - two brane models
with $\Omega_{\lambda,0} \ne 0$: lower - the best-fit
model; higher - the best-fit {\it flat} model.
The brane phantom plots are very close to the top line of Ref.
\cite{Perlmutter99}.}
\label{Fig.2}
\end{figure}

In Fig. \ref{Fig.2} we present the difference in redshift-magnitude relation
(\ref{chir1}) for brane models with phantom matter on the brane ($\gamma =
-1/3$) using Perlmutter's sample A. Note that the theoretical curves are very close to that of
\cite{Perlmutter99} which means that the phantom
cancels the positive-pressure influence of the $\varrho^2$ term
and can {\it mimic} the negative-pressure influence of the cosmological constant to cause
cosmic acceleration. From the formal point of view the best fit
is ($\chi^2=95.4$) for $\Omega_{k,0}= 0.2$, $\Omega_{ph,0}=0.7$, $\Omega_{\lambda,0}=-0.1$,
$\Omega_{\cal U} = 0.2$, $\Omega_{\Lambda_{4},0}=0$ which
means that the cosmological constant must necessarily {\it vanish}. From
this result we can conclude that phantom matter $p = - (4/3) \varrho$ can {\it
mimic} the contribution from the $\Lambda_{4}$-term in standard models.
For the best-fit flat model ($\Omega_{k,0}=0$) we have
($\chi^2=95.4$): $\Omega_{ph,0}=0.2$, $\Omega_{\lambda,0}=-0.1$,
$\Omega_{\cal U} = 0.2$, $\Omega_{\Lambda_{4},0}=0.7$.

\section{Angular diameter size and the age of the universe}

Another cosmological test which can be performed is the angular diameter
size
\be
\theta = \frac{d(z+1)^2}{d_{L}}
\ee
of a galaxy with a linear size $d$. In a flat dust ($\gamma
=1$) universe $\theta$ has the minimum value $z_{min} = 5/4$.
It is particularly interesting to notice that
for flat brane models with $\Omega_{\lambda} \approx 0, \Omega_{\Lambda_{4}}
\approx 0$ the dark radiation can {\it enlarge} the minimum value of $\theta$
while the ordinary radiation lowers this value \cite{AJI+II}, i.e.,
\be
\label{zminU}
z_{min} = \frac{1}{2{\cal U}} \left( \Omega_{\cal U} - 1 + \sqrt{3
\Omega_{\cal U} + 1} \right) \ge \frac{5}{4}
\ee
for $\Omega_{\cal U} \leq 0$. This is a general influence of negative dark radiation
onto the angular diameter size for brane models. One can also notice that there
exists a restriction on the amount of negative dark radiation coming from
(\ref{zminU}) $(\Omega_{\cal U} \geq - 1/3)$ which can serve
as a test for the admissible value of $\Omega_{\cal U} = - 1/3$ ($z_{min}
=2$) in order to observe the minimum.

\begin{figure}[h]
\includegraphics[scale=.28]{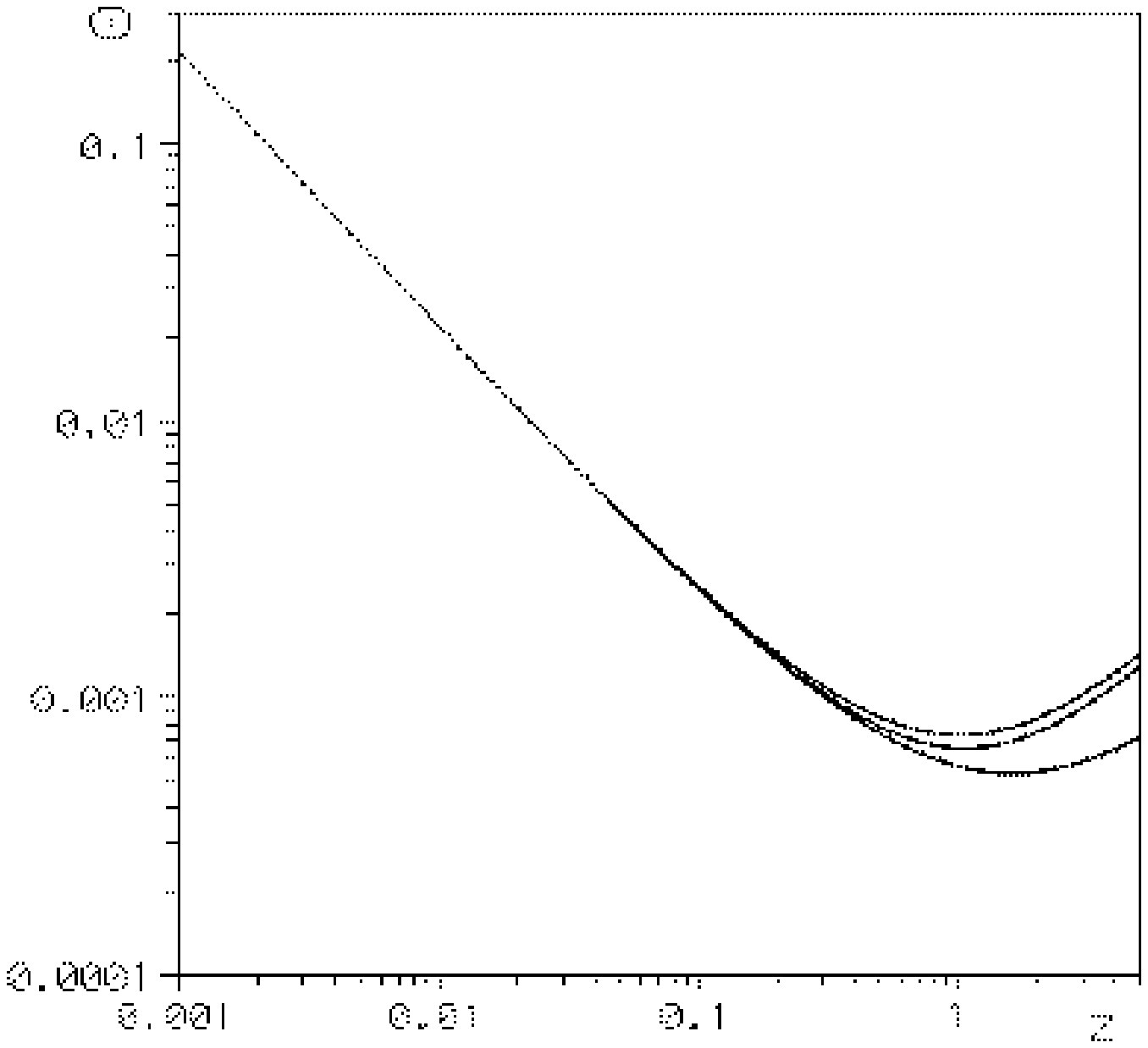}
\caption{The angular diameter $\theta$ for $\Omega_{\lambda} = 0.1, \Omega_m = 0.3,
\Omega_{\Lambda_{4}} = 0.72$, and the two values of $\Omega_{\cal U} = 0.1, -0.1$
(top, middle) in comparison with the model of Ref. \cite{Perlmutter99} with $\Omega_m = 0.28,
\Omega_{\Lambda_{4}} = 0.72$ (bottom).}
\label{Fig.5}
\end{figure}

More detailed analytic and numerical studies show
that the increase of $z_{min}$ is even more sensitive to negative
values of $\Omega_{\lambda}$ which, unfortunately,
are admissible only for timelike extra dimensions \cite{Shtanov}. Similarly as for the dark
radiation $\Omega_{\cal U}$, the minimum disappears for some large negative $\Omega_{\lambda}$.
Positive $\Omega_{\cal U}$ and $\Omega_{\lambda}$ make $z_{min}$ decrease.
In Fig. \ref{Fig.5} we present a plot from which one can see the
sensitivity of $z_{min}$ to $\Omega_{\cal U}$.

\begin{figure}
\includegraphics[width=0.28\textwidth]{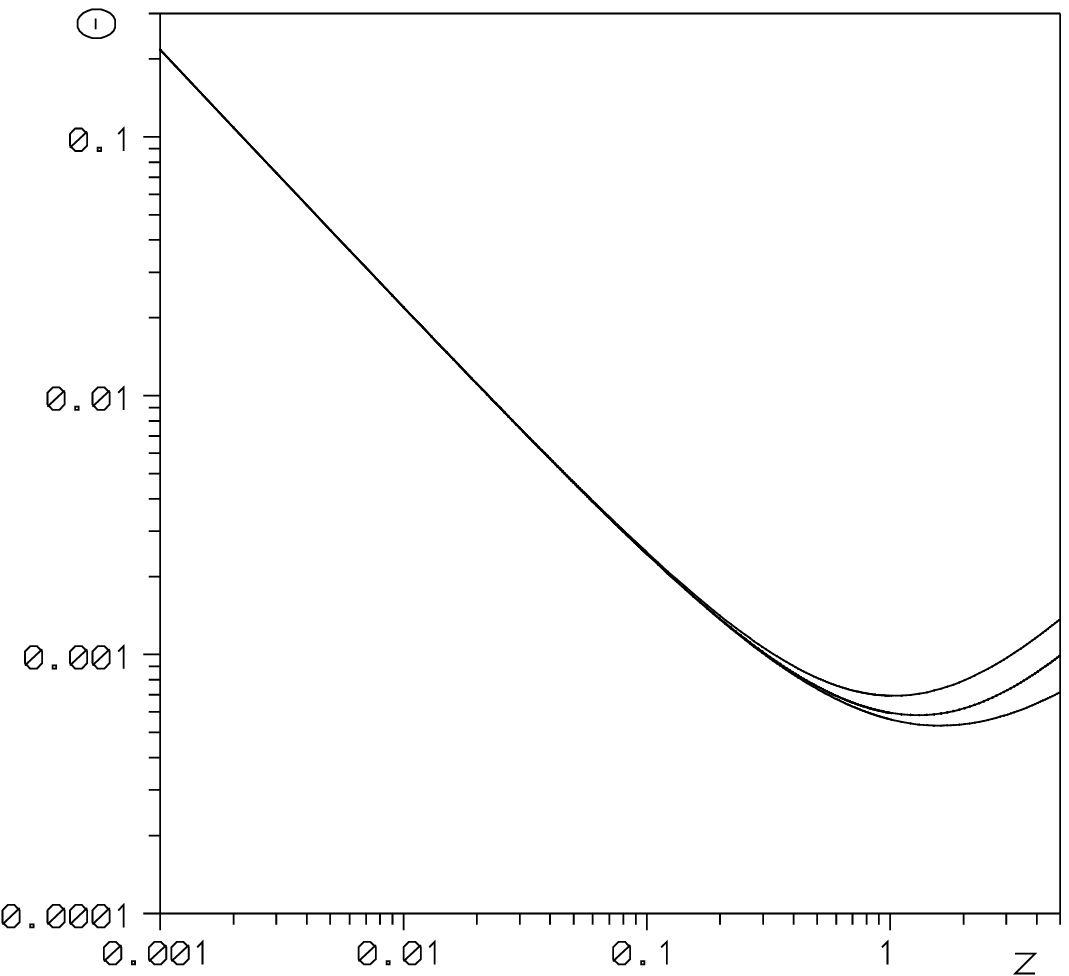}
\caption{The angular diameter $\theta$ for the flat brane model for
$\Omega_{\mathrm{m},0} = 0.3$ and $\Omega_{\lambda,0} = 0.1,0.02,0$
(top, middle, bottom, respectively). The minima for these cases are $1.04$,
$1.30$, $1.605$, respectively. The brane fluid causes the minimum to move
left (towards to lower $z$) and the minimum value of $\theta$ increases.}
\label{fig:11}
\end{figure}

As far as a possible contribution of $\Omega_{\lambda}$ is concerened, then
from Fig.~\ref{fig:11} we can see that
its influence on the angular size $\theta$ is relatively weak.
As a conclusion we can say that it is possible
to test values of $\Omega_{{\cal U},0}$ and $\Omega_{\lambda,0}$ from the angular
diameter minimum value test, but because of the evolutionary effects and observational
difficulties the predicted differences are too small to be detect.
We have also checked that phantom matter $\Omega_ph$ has very little influence
onto the value of $z_{min}$.

\begin{figure}[h]
\includegraphics[angle=270,scale=.28]{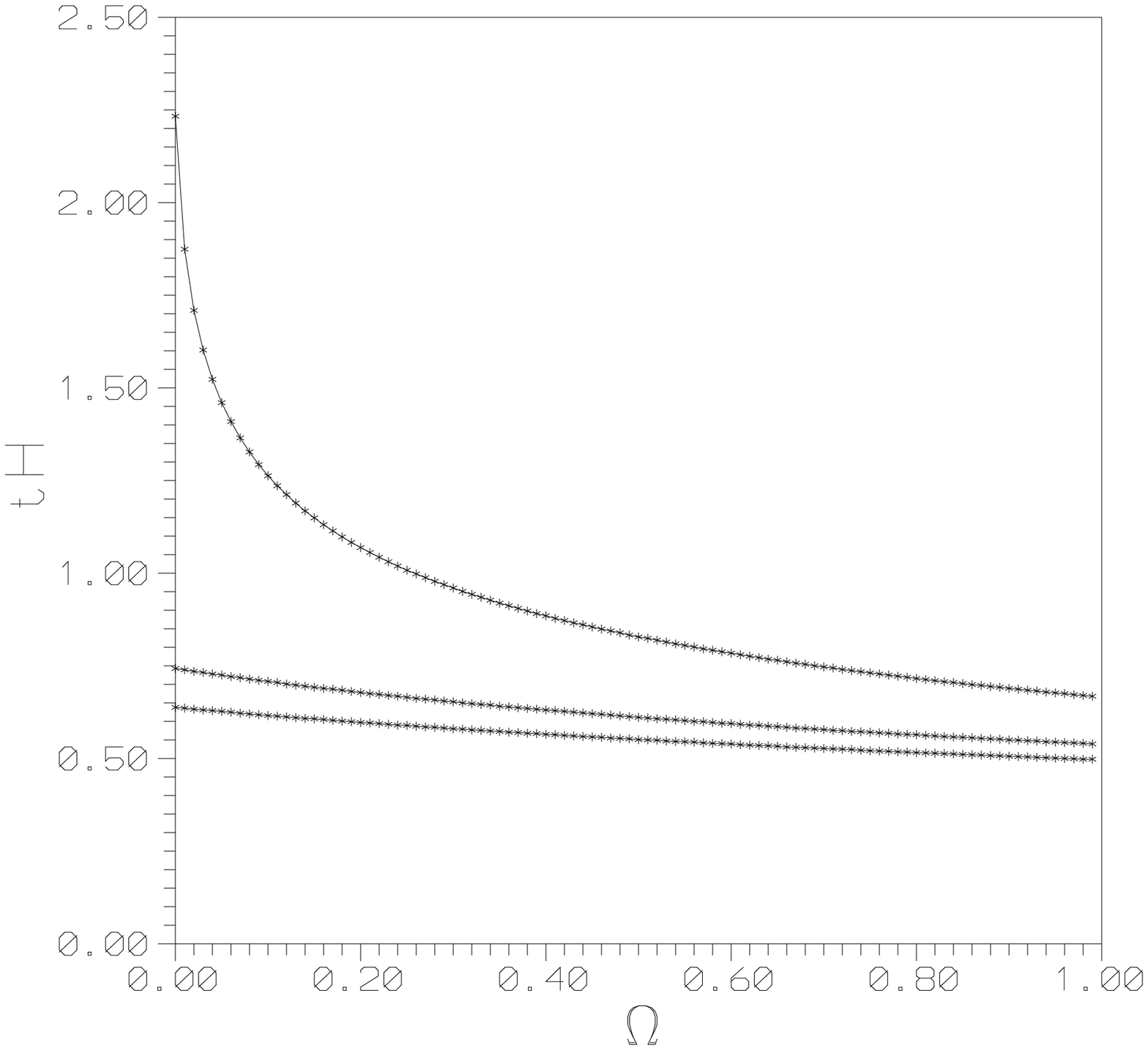}
\caption{The age of the universe $t_0$ in units of $H_0^{-1}$ for the brane models with dust ($0 \leq \Omega_{m,0} \leq 1$
on the horizontal axis). Here $\Omega_{{\cal U},0} = \Omega_{k,0} = 0$,
$\Omega_{{\lambda},0} = 0, 0.05, 0.1$ (top, middle, bottom).}
\label{Fig.3}
\end{figure}

Now let us briefly discuss the effect of brane parameters and phantom matter onto
the age of the universe which according to (\ref{FriedCCC}) is given by
\bea
\label{age}
H_0 t_0 = \int_0^1 \left\{ \Omega_{GR,0} x^{-3\gamma +
4} + \Omega_{{\lambda},0} x^{-6\gamma + 4}
\right. \nonumber \\ \left.
+ \Omega_{{\cal U},0} +
\Omega_{k,0} x^2 + \Omega_{\Lambda_{4},0} x^4 \right\}^{-\frac{1}{2}} x dx  ,
\eea
where $x = a/a_0$. We made a plot for the dust $\gamma = 1$ on the brane in
Fig.\ref{Fig.3} which shows that the effect of quadratic in
energy density term represented by $\Omega_{\lambda}$ is to {\it
lower} significantly the age of the universe.

For the dust matter and the cosmological constant the age of universe
is given by
\be
\label{t0}
t_{0} = \frac{1}{3 H_{0} \sqrt{\Omega_{\Lambda,0}}}
\ln \frac{\Omega_{\mathrm{m},0} + 2\Omega_{\Lambda,0}
+ 2\sqrt{\Omega_{\Lambda,0}}}{\Omega_{\mathrm{m},0}
+ \sqrt{4\Omega_{\lambda,0}\Omega_{\Lambda,0}}}.
\ee
In Fig.~\ref{fig:12} we made another plot of (\ref{t0}) in Gyrs for the
flat model for different $\Omega_{\lambda,0}$.
We can again see that the dust matter on the brane which mimics brane contribution
as stiff-fluid decreases significantly the age of the
universe.

\begin{figure}[h]
\includegraphics[angle=270,scale=.28]{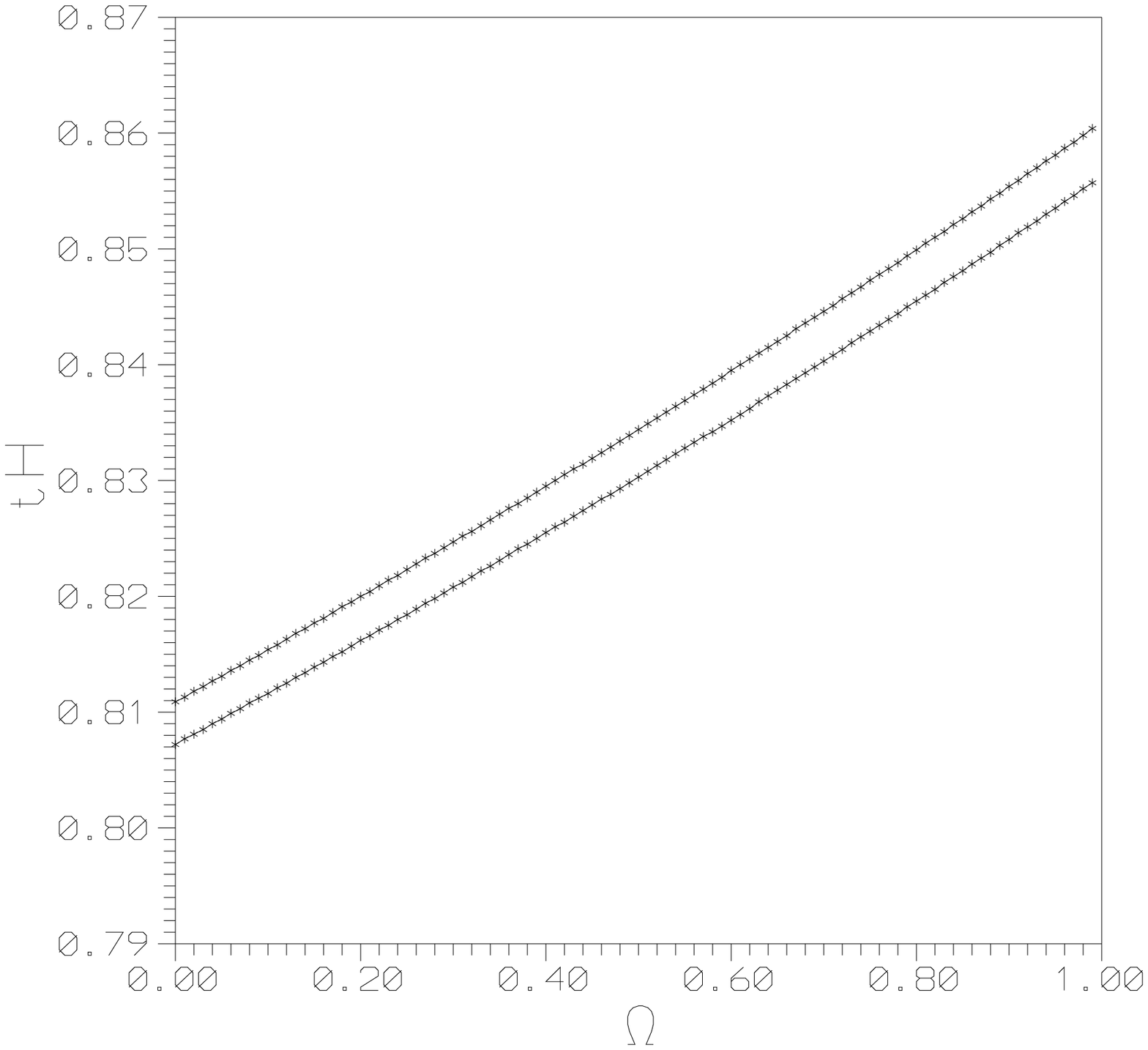}
\caption{The age of the universe $t_0$ in units of $H_0^{-1}$ for the brane models with
phantom on the brane
($0 \leq \Omega_{ph,0} \leq 1$ on the horizontal axis). Here $\Omega_{{\cal U},0} =
0.2$, $\Omega_{{\lambda},0} = 0.05, 0$ (top, bottom) which shows weaker influence
of the brane effects to increase the age.}
\label{Fig.4}
\end{figure}

\begin{figure}
\includegraphics[width=0.28\textwidth]{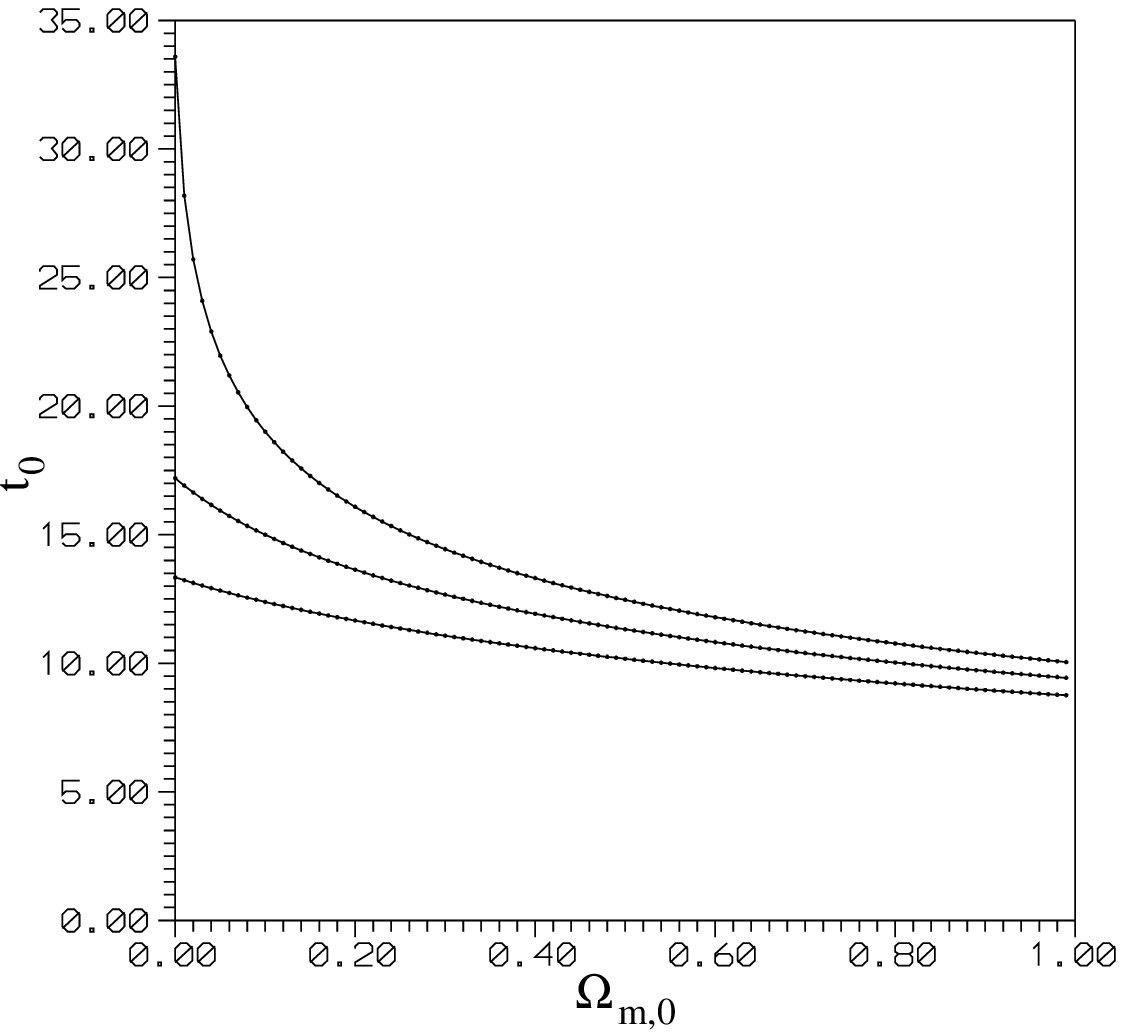}
\caption{The age of the universe $t_{0}$ in units of $10^9$ yrs for
the flat brane model with $\Omega_{\lambda,0} = 0, 0.004, 0.020$
(top, middle, bottom).}
\label{fig:12}
\end{figure}

The problem can be
avoided, if we {\it accept} phantom $\gamma = - 1/3$
on the brane \cite{darkenergy}, since the phantom
has a very strong influence to increase the age.
In Fig. \ref{Fig.4} we made a plot for this case which shows how
phantom energy enlarges the age.

Therefore the {\it cosmological constant} or the {\it phantom matter} is needed to
explain the problem of the age of the universe.

\section{Doppler peaks}

The CMB peaks arise from acoustic oscillations of the primeval plasma.
Physically, these oscillations are represented by hot and cold spots in CMB
temperature. A wave
which had a maximum of the density at the time of the last scattering corresponds to
a peak in the power spectrum. In the Legendre multipole space this
corresponds to an appropriate angle $\theta_a$ subtended by the sound horizon at the last
scattering surface. Higher harmonics of the principal oscillations, which have
oscillated more than once, correspond to secondary peaks.

For our purpose it is very important that the location of these peaks are
very sensitive to the variations in the model parameters. Therefore, it can be
used as a probe to constrain the cosmological parameters and
discriminate among various models.

If ${\theta}_a$ is the angular size of the density fluctuation, then the
acoustic scale for the peaks is ($l_a=\pi/\theta_a$) \cite{peaks}
\be
l_a=\pi \frac{\int\limits_0^{z_{dec}} [f(z^{'})]
^{-1/2}dz^{'}}{\int\limits_{z_{dec}}^{\infty} c_s [f(z^{'})]
^{-1/2}dz^{'}}
\ee
($z_{dec}=1100$) where $c_s$ is the speed of sound and the position of the
peaks is
\be
\label{el}
l_{i}=l_a(i-\phi_i)
\ee
(i = 1,2,3), where the phase shift \cite{Doran01,Hu01}
\be
\phi_{i} \approx 0.267 \left[ \frac{r(z_{\mathrm{dec}})}{0.3}
\right]^{0.1}.
\ee

The phase shift is caused by the pre-recombination physics (plasma driving
effect) and, hence, has no significant contribution from the term containing
brane in that epoch so that it was taken from standard cosmology.
We take $\Omega_{\mathrm{b},0} h^{2} = 0.02$,
$r(z_{\mathrm{dec}}) \equiv
\rho_{\mathrm{r}}(z_{\mathrm{dec}})/\rho_{\mathrm{m}}(z_{\mathrm{dec}})
= \Omega_{\mathrm{r},0}(1 + z_{\mathrm{dec}})/\Omega_{\mathrm{m},0}$,
$\Omega_{\mathrm{r},0} = \Omega_{\gamma,0} + \Omega_{\nu,0}$,
$\Omega_{\gamma,0} = 2.48h^{-2} \cdot 10^{-5}$,
$\Omega_{\nu,0} = 1.7h^{-2} \cdot 10^{-5}$, $r(z_{\mathrm{dec}})$
is the ratio of radiation to matter densities at the surface of last
scattering and we have
$\Omega_{\mathrm{r},0} = 9.89 \cdot 10^{-5}$,
$\Omega_{\mathrm{b},0} = 0.05$, and the spectral index for initial
density perturbations $n = 1$, and $h = 0.65$
\cite{Parampreet02}.

For dust matter $\Omega_{m,0}$, radiation $\Omega_{r,0}$, and dark radiation
$\Omega_{{\cal U},0}$ ($\Omega_{{\gamma},0}= \Omega_{r,0} + \Omega_{{\cal
U},0}$) on the brane under the assumption that we neglect the influence of the
quadratic in energy density term of radiation $\Omega_{r,0}$ we have the effective
speed of sound
\be
c_{eff}^2= c_s^2+
{\Omega_{b,0}\left(1+z \right)^3+{4\over 3}\Omega_{\gamma,0}\left(1+z \right)^4
\over \Omega_{\tilde{\lambda},0}+
\Omega_{b,0}\left(1+z \right)^3+\Omega_{\gamma,0}\left(1+z
\right)^4},
\ee
with $c_s^2$ of the same form as in standard cosmology and
\be
\Omega_{\tilde{\lambda},0} =
\frac{\Omega^2_{m,0}}{2\Omega_{\lambda,0}} ~.
\ee

The influence of brane on the location of the peaks is to shift them
towards higher values of $l$. For example, for $\Omega_{\mathrm{m},0} = 0.3$,
$\Omega_{\mathrm{b},0} = 0.05$, $h = 0.65$, the different choices of
$\Omega_{\lambda,0}$ yield the following
\begin{eqnarray*}
\Omega_{\lambda,0} &=& 0 \colon  l_{\mathrm{peak},1} = 225,
 l_{\mathrm{peak},2} = 535,  l_{\mathrm{peak},3} = 847\\
\Omega_{\lambda,0} &=& 1.5 \cdot 10^{-15} \colon  l_{\mathrm{peak},1} =
227,
 l_{\mathrm{peak},2} = 540,  l_{\mathrm{peak},3} = 853,\\
\Omega_{\lambda,0} &=& 10^{-12} \colon  l_{\mathrm{peak},1} =
239,
l_{\mathrm{peak},2} = 568,  l_{\mathrm{peak},3} = 897.
\end{eqnarray*}

We could also analyze the influence of dark radiation term in
the brane world cosmology. The corresponding term in this case scales
just like radiation with a constant $\rho_0$ and both positive and negative
values of $\rho_{\mathrm{r},0}$ ($\rho_{{\cal U},0}$) are possible.
Dark radiation should strongly affect both the BBN and CMB. Ichiki {\em et al.\/}
\cite{Ichiki02} used such observations to constrain both the magnitude
and the sign of dark radiation in the case when the $\rho^2$ term  coming
from the brane is negligible.
We take their limits on dark radiation with negative contribution
because of the tension
between the observed ${}^4\mathrm{He}$ and $D$ abundance \cite{Ichiki02}.
In our case we take into consideration the brane term $\rho^2$
and obtain the following positions of the first three peaks:
\begin{tabular}{lll}
$\Omega_{\mathrm{\cal U},0} = -1.23\Omega_{\gamma,0}$
& $\Omega_{\lambda,0} = 10^{-12} \colon$ & $l_{\mathrm{peak},1} = 208$,
$l_{\mathrm{peak},2} = 495$, \nonumber \\  $l_{\mathrm{peak},3} = 781$, \\
$\Omega_{\mathrm{\cal U},0} = -\Omega_{\gamma,0}$
& $\Omega_{\lambda,0} = 10^{-12} \colon$ & $l_{\mathrm{peak},1} = 214$,
$l_{\mathrm{peak},2} = 545$, \nonumber \\ $l_{\mathrm{peak},3} = 861$, \\
$\Omega_{\mathrm{\cal U},0} = -0.41\Omega_{\gamma,0}$
& $\Omega_{\lambda,0} = 10^{-12} \colon$ & $l_{\mathrm{peak},1} = 229$
$l_{\mathrm{peak},2} = 545$, \nonumber \\ $l_{\mathrm{peak},3} = 861$.
\end{tabular}

From Boomerang observations \cite{deBernardis02}
we obtain $l_{\mathrm{peak},1} = 200-223$, $l_{\mathrm{peak},2} = 509-561$.
Therefore, uncertainties in values $l_{\mathrm{peak}}$ can be used in
constraining cosmology with brane effect, namely
\[
\Omega_{\lambda,0} \leq 1.0 \cdot 10^{-12}
\]
from the location of the first peak.

We also compare the results from the above procedure with recent bounds on the
location of the first two peaks obtained by WMAP
collaboration \cite{Spergel03,Page03} namely
$l_{\mathrm{peak},1} = 220.1 \pm 0.8$, $l_{\mathrm{peak},2} = 546 \pm 10$,
together with the bound on the location of the third peak obtained by
Boomerang collaborations $l_{\mathrm{peak},3} = 825^{+10}_{-13}$ which lead to
quite strong constraints on the model parameters. These constraints can
be summarized as follows. If we assume no dark radiation, the brane model
is in agreement with observations provided we take
\[
\Omega_{\lambda,0} \leq  1.0 \cdot 10^{-15}
\]

We have made similar considerations for phantom
on the brane (with matter and radiation included - otherwise the
speed of sound becomes imaginary, compare Ref. \cite{freese02}) and obtained the limit
$\Omega_{{\lambda},0} \le 1.0 \times 10^{-13}$ for $\Omega_m=0.3$,
$\Omega_{ph,0} =0.7-\Omega_{{\gamma},0}, \Omega_{k,0} = 0$ and the
location of the peaks is at $l_1=225, l_2=535, l_3=845$.

However, the phase shift $\phi$ in the above considerations
is taken from the standard cosmology,
i.e., we assume that the contribution from the brane term is insignificant
at the pre-recombination epoch. If this assumption was not valid then the
limit from CMB would change.

\section{Big-Bang nucleosynthesis}

It is clear from the Friedmann equation (\ref{Friedroro}) that the
brane models with positive pressure $\gamma > 1$ matter lead to a
dominance of the $\varrho^2$ term in the early universe. However,
its admittance, even for the late universe can be dominant. In
particular, if the restriction we found for dust matter on the brane from supernovae
$\Omega_{\lambda ,0} \sim 0.01$ is to be applied, then the brane term
can dominate the universe already at redshift $z \sim 2$.

For large $z$ the brane term which comes from radiation scales like $(1+z)^8$ and
it dominates over the standard radiation term. This brings a potential trouble
since in such a model radiation domination never occurs and
all BBN predictions fail. In other words, the preferred value
obtained from SN Ia data gives the $\rho^{2}$ term which is
far too large to be compatible with BBN, if we assume that brane models
do not change the physics in the pre-recombination epoch.

The consistency with BBN seems to be
a crucial issue in brane cosmology \cite{Arkani-Hamed99,Binetruy00,Bratt02}.
For this reason, we should admit that
the contribution of brane fluid $\Omega_{\lambda,0}$ cannot dominate over
the standard radiation and dark radiation terms before the onset of BBN,
i.e., for $z \cong 10^8$ which gives a condition
\be
\label{BBN}
\Omega_{\lambda,0}\frac{\Omega^2_{\mathrm{r},0}}{\Omega^2_{\mathrm{m},0}}(1+z)^8
< (\Omega_{\mathrm{r},0}+ \Omega_{\mathrm{\cal U},0})(1+z)^4
\ee
where $\Omega_{\mathrm{\cal U},0} \leq 0.11 \Omega_{\mathrm{r},0}$ \cite{Ichiki02}.

Therefore, the term
\be
\Omega_{\lambda,0}\frac{\Omega^2_{\mathrm{r},0}}{\Omega^2_{\mathrm{m},0}}(1+z)^8,
\ee
which describes brane effects
is constrained by the BBN because it requires a change of the expansion rate
due to this term to be sufficiently small, so that an acceptable helium-4
abundance is produced. Taking this into account we obtain the following limit in this case
\[
\Omega_{\lambda,0} \leq 1.0 \cdot 10^{-27} \quad \mathrm{if} \quad z \simeq 10^8.
\]

However, the situation can significantly be changed if we admit
{\it phantom matter} in the universe because the phantom brane term
never dominates at large redshifts. Quite the contrary, it eventually
dominates present and future evolution and causes no change of BBN
predictions. This is because it effectively leads to a reversal of eq. (\ref{BBN})
in a similar way as it reverses the luminosity distance relation
in eq. (\ref{phandL}).

\section{Conclusions}

We have investigated observational constraints for the Randall-Sundrum scenario
of brane world cosmology. The motivation to study
new observational constraints for some particular scenarios of the brane world
cosmology is the demonstration that such models inspired by recent developments
in the particle physics can be tested by astronomical observations.

We have shown that this scenario is compatible with the most recent observations
of SN Ia. Moreover, we demonstrate that only a new (high $z$) and more
precise set of observations can show whether the considered class of models
constitute a viable possibility for the description of the present acceleration
of the Universe.

The brane model with dust $\gamma=1$ matter on the brane fits the supernovae Ia data
very well (including SN 1997ff at $z \simeq 1.7$). However, the cosmological constant
is still required.
We obtain the best-fit non-flat model with $\Omega_{\lambda,0} \simeq 0.01$,
$\Omega_{k,0} \simeq -0.9$, $\Omega_{\mathrm{m},0} = 0.6$, $\Omega_{\Lambda,0}
=1.3$.
For the flat model with $\Omega_{\mathrm{m},0}=0.3$ we obtain
$\Omega_{\lambda_,0}=0.004 \pm 0.016$ as a best-fit. Whereas the best-fit
non-flat model is not realistic (because of large negative curvature),
the flat model with the brane for
the realistic value of $\Omega_{\mathrm{m},0}=0.3$ is in agreement with
the SN Ia data.

On the other hand, brane models with phantom $\gamma = -1/3$
matter on the brane also fit supernovae data and they may {\it mimic} the
contribution from the cosmological constant which is then not required.

We have demonstrated how the values of the estimated parameters depend on the division
of the sample on high and low redshifts subsamples. As a results we obtained that although
the full data set (Tonry/Barris sample) of SNIa require the cosmological
term, the low redshifts $z \le 0.25$ data set, admits
decelerating model without cosmological constant. We have also
demonstrated
the sensitivity of results with respect to small changes of ${\cal M}$ parameter.

It is interesting that brane models with dust $\gamma=1$ matter for high
redshifts predict brighter galaxies than the Perlmutter model. Therefore, the
difference between
the Perlmutter model and the brane model may be detectable for high
redshit $z > 1.2$ supernovae.

Let us note that present data suggest that $\Omega_{\lambda,0} 1.0 \simeq 10^{-2}$
but because of the large error of this estimation it is possible that
$\Omega_{\lambda,0}=0$. However, from future observations of supernovae
at high redshift we can expect that errors in estimation of $\Omega_{\lambda,0}$
become significantly smaller. At present the brane theory can be neither
confirmed nor ruled out. In this way the future SN Ia data should allow
to verify the hypothesis that $\Omega_{\lambda,0}$ is so large as
$\Omega_{\lambda,0}\simeq 0.01$.

We also found the other limits on the value of $\Omega_{\lambda,0}$ from
the measurements of CMB anisotropies and BBN. We obtain the strongest limits in
these cases, namely $\Omega_{\lambda,0} < 1.0 \cdot 10^{-12}$ from CMB
and $\Omega_{\lambda,0} \le 1.0 \cdot 10^{-27}$ from BBN.  However, let us note that
because the errors in estimation of $\Omega_{\lambda,0}$ from supernovae
are so large the results obtained from SN Ia do not contradict those of
BBN and CMB.

Of course BBN as well as CMB are very well tested areas of cosmology which
do not allow for a significant deviation from the standard model and standard
expansion law, except at very early times. Although consistency with
BBN and CMB is a crucial issue in brane models, we must remember that
in such an approach we assume that brane models does not change the physics
in the pre-recombination epochs.
On the other hand, the results of analysis SNIa data are independent of the physical
processes in the early universe. Therefore weaker limits obtained
from SNIa  observations may even be more valuable.

\section{Acknowledgements}

We wish to thank Dr. B.J. Barris for explanation the details of his SNIa sample.
The support from Polish Research Committee (KBN) grants No 2P03B 090 23 (M.P.D.)
and No 2P03B 107 22 (M.S.) is acknowledged.


\end{document}